\documentclass[a4paper,11pt]{article}
\pdfoutput=1 % if your are submitting a pdflatex (i.e. if you have
\usepackage{jcappub} % for details on the use of the package, please
\renewcommand{\v}[1]{\ensuremath{\mathbf{#1}}} % for vectors
\newcommand{\gv}[1]{\ensuremath{\mbox{\boldmath$ #1 $}}} 
\newcommand{\grad}[1]{\gv{\nabla} #1} % for gradient
\let\baraccent=\= % rename builtin command \= to \baraccent
\renewcommand{\=}[1]{\stackrel{#1}{=}} % for putting numbers above =

\usepackage{graphicx}% Include figure files
\usepackage{dcolumn}% Align table columns on decimal point
\usepackage{multirow}
\usepackage{natbib}
\usepackage{xcolor}
\usepackage{mathtools}
\usepackage{hyperref}
\usepackage[normalem]{ulem}
\usepackage{floatrow}
\newfloatcommand{capbtabbox}{table}[][\FBwidth]

\title{X-Ray Shapes of Elliptical Galaxies and Implications for Self-Interacting Dark Matter}

\author[a,bc,1]{A. McDaniel,\note{Corresponding author.}}
\author[b,c]{T., Jeltema,}
\author[b,c]{S., Profumo}

\affiliation[a]{Department of Physics and Astronomy, Kinard Lab of Physics, Clemson, SC, 29634-0978, USA}
\affiliation[b]{Department of Physics, University of California,
1156 High Street, Santa Cruz, California, 95064, USA}
\affiliation[c]{Santa Cruz Institute for Particle Physics,
1156 High Street, Santa Cruz, California, 95064, USA}

\emailAdd{armcdan@clemson.edu}
\emailAdd{tesla@ucsc.edu}
\emailAdd{profumo@ucsc.edu}

\abstract{Several proposed models for dark matter posit the existence of self-interaction processes that can impact the shape of dark matter halos, making them more spherical than the ellipsoidal halos of collisionless dark matter. One method of probing the halo shapes, and thus the strength of the dark matter self-interaction, is by measuring the shape of the X-ray gas that traces the gravitational potential in relaxed elliptical galaxies. In this work we identify a sample of 11 relaxed, isolated elliptical galaxies and measure the ellipticity of the gravitating matter using  X-ray images from the XMM-Newton and Chandra telescopes. We explore a variety of different mass configurations and find that the dark matter halos of these galaxies have ellipticities around $\epsilon\approx 0.2-0.5$. While we find non-negligible scatter in the ellipticity distribution, our results are consistent with some degree of self-interaction at the scale of $\sigma/m \sim 1$ cm$^2$/g, yet they also remain compatible with a  cold dark matter scenario. We additionally demonstrate how our results can be used to directly constrain specific dark matter models and discuss  implications for current and future simulations of self-interacting dark matter models.}

\begin{document}

%%%% MACROS %%%%
%\let\jnl@style=\rm
\def\reff@jnl#1{{\rm#1}} %NOTE: removed \jnl@style, replaced with \rm

\def\araa{\reff@jnl{ARA\&A}}             % Annual Review of Astron and Astrophys
\def\aj{\reff@jnl{AJ}}                   % Astronomical Journal
\def\apj{\reff@jnl{ApJ}}                 % Astrophysical Journal
\def\apjl{\reff@jnl{ApJ}}                % Astrophysical Journal, Letters
\def\apjs{\reff@jnl{ApJS}}               % Astrophysical Journal, Supplement
\def\apss{\reff@jnl{Ap\&SS}}             % Astrophysics and Space Science
\def\aap{\reff@jnl{A\&A}}                % Astronomy and Astrophysics
\def\aapr{\reff@jnl{A\&A~Rev.}}          % Astronomy and Astrophysics Reviews
\def\aaps{\reff@jnl{A\&AS}}              % Astronomy and Astrophysics, Supplement
\def\baas{\ref@jnl{BAAS}}               % Bulletin of the AAS
\def\jcap{\reff@jnl{J. Cosmology Astropart. Phys.}}% JCAP
\def\jrasc{\reff@jnl{JRASC}}             % Journal of the RAS of Canada
\def\memras{\reff@jnl{MmRAS}}            % Memoirs of the RAS
\def\mnras{\reff@jnl{MNRAS}}             % Monthly Notices of the RAS
\def\na{\reff@jnl{New A}}                % New Astronomy
\def\nar{\reff@jnl{New A Rev.}}          % New Astronomy Review
\def\pra{\reff@jnl{Phys.~Rev.~A}}        % Physical Review A: General Physics
\def\prb{\reff@jnl{Phys.~Rev.~B}}        % Physical Review B: Solid State
\def\prc{\reff@jnl{Phys.~Rev.~C}}        % Physical Review C
\def\prd{\reff@jnl{Phys.~Rev.~D}}        % Physical Review D
\def\pre{\reff@jnl{Phys.~Rev.~E}}        % Physical Review E
\def\prl{\reff@jnl{Phys.~Rev.~Lett.}}    % Physical Review Letters
\def\pasa{\reff@jnl{PASA}}               % Publications of the Astron. Soc. of Australia
\def\pasp{\reff@jnl{PASP}}               % Publications of the ASP
\def\pasj{\reff@jnl{PASJ}}               % Publications of the ASJ
\def\qjras{\reff@jnl{QJRAS}}             % Quarterly Journal of the RAS
\def\physrep{\reff@jnl{Phys.~Rep.}}   % Physics Reports
%TESTING \nat
\def\nat{\reff@jnl{Nature}}   % Physics Reports
\let\astap=\aap
\let\apjlett=\apjl
\let\apjsupp=\apjs
\let\applopt=\ao
%%%% END MACROS %%%%

\maketitle
\flushbottom
\section{\label{sec:intro}Introduction}
The cold dark matter (CDM) paradigm has been immensely successful in explaining many aspects of the universe and is particularly successful at describing large-scale structure \cite{LSS}. This has naturally led to CDM as the benchmark for dark matter (DM) studies and the most extensively investigated class of DM candidates. However, the lack of indirect or direct observational evidence of collisionless cold dark matter (e.g. WIMPs) \cite{Arcadi:2017kky}, along with purported shortcomings of CDM at small scales \cite{KlypinMS, mooreMS,TBTF2011, TBTF2012,FloresPrimack, mooreCC, KamadaCC} has led to interest in alternative DM frameworks. In regards to the latter point, challenges at small scales such as the missing satellites \cite{KlypinMS, mooreMS}, ``Too-Big-To Fail'' \cite{TBTF2011, TBTF2012},  and core-cusp \cite{FloresPrimack, mooreCC, KamadaCC} problems are potentially troubling for the CDM paradigm (see however \cite{primackNature,navarroDwarfs,governatoCusps,brooksDwarfs, ZolotovBrooks, bullockReview}). Solutions to these problems based on baryonic effects alone have been explored \cite{brooksDwarfs, navarroDwarfs, governatoDwarfs, amorisco}, though it is not clear that these effects are sufficient to account for the discrepancies between observations and simulations \cite{kirby2014,kapKeeleyLinden2014,PapPon2017}. An intriguing solution to these problems is to consider self-interacting dark matter (SIDM) that is not fully collisionless, which was first proposed in \cite{SS} and recently reviewed in \cite{tulinYuReview}.

N-body simulations as well as analytical arguments suggest that DM self-interactions can yield observable effects on the macroscopic properties of halos that can impact and potentially even address some of the problems that arise in a collisionless CDM scenario. Specifically, self-interactions can flatten the centrally peaked cusps in the inner regions of galaxies and are capable of disrupting the development of the dense satellite galaxies expected from simulations that are at the crux of the core-cusp and TBTF problems \cite{VZL2012, Peter2013, Elbert2015}. SIDM is also predicted to affect the shapes of DM halos through isotropized particle scattering \cite{Peter2013, SS,DSS2000}, producing more spheroidal halos than seen in CDM. These predictions for the macroscopic effects of SIDM provide an opportunity to probe the microscopic properties of DM-DM interactions. The figure of merit for SIDM is the ratio of the collisional cross-section to the DM mass $\sigma/m$ , since the scattering rate for DM self-interactions is proportional to the {\em number} density of dark matter particles $n_{\rm DM}=\rho_{\rm DM}/m$ times the thermally averaged self-interaction cross section times the relative velocity, $\langle \sigma v_{\rm rel}\rangle$, thus $\Gamma_{\rm DM-DM}=n_{\rm DM}\langle \sigma v_{\rm rel}\rangle$. In order to alleviate the small-scale challenges discussed above, several groups have found that the necessary interaction strength is roughly $\sigma/m \approx 0.5-10$ cm$^2$/g  \cite{VZL2012, Peter2013, Elbert2015,SS,DSS2000, bullockReview}. This is in line with the simple expectation that for typical relative DM velocities the rate $\Gamma_{\rm DM-DM}\sim H$, with $H$ the Hubble rate.

Studies of the predicted effects of SIDM in comparison with observations place some constraints on the ratio of the DM self-interaction cross section to the dark matter mass. These observational probes include cluster lensing \cite{ME02,Peter2013, andrade}, mergers \cite{randallBullet,M04Bullet}, and X-ray ellipticities \cite{Buote02}. While early constraints on SIDM in clusters suggested low cross section values of $\sigma/m\lesssim 0.02$ cm$^2$/g \cite{ME02}, subsequent higher resolution simulations indicated that these limits were overly optimistic and that cross sections of $\sigma/m  \lesssim 1$ cm$^2$/g are consistent with simulations across a variety of mass scales \cite{Rocha,Peter2013}.  Furthermore, observations at a wide range of halo masses indicate that a velocity dependent cross-section is needed in order to alleviate small-scale issues while also being consistent with cluster constraints \cite{SIDMVelocity}, thus providing compelling motivation for studies at a range of halo mass scales. Specifically, the interaction cross-section must decrease with larger relative velocity (e.g in clusters). This can naturally be achieved in SIDM models wherein the interactions are governed by a Yukawa potential (see for instance the models described in Ref.~\cite{fengKap,loeb,buckley,TYZ1,TYZ2}.

An interesting class of targets for SIDM studies is elliptical galaxies with halo masses of order $\sim 10^{12}-10^{13}M_{\odot}$. Elliptical galaxies are interesting to DM studies in part because the interstellar X-ray emitting gas fills the gravitational potential, acting as a tracer of the underlying gravitational potential, well beyond the range where stellar dynamics can be used as a probe of the mass content of the system \cite{BH_review}. However, this relies on the assumption that the gas is in a state of hydrostatic equilibrium with the potential. If the elliptical galaxies in question can reasonably be treated as being in hydrostatic equilibrium, the gas traces the gravitating potential, and determination of the shape of the X-ray gas allows for probes of the DM dominated mass distribution. Thus, by observing the X-ray shapes of {\em relaxed} (i.e. in thermal equilibrium) elliptical galaxies, the shape of the DM halo can be inferred and compared with predictions of CDM and SIDM halos shapes from simulations. This ``X-ray Shape Theorem'' first was developed by \cite{BinneyStrimpel} and has been applied to studies of elliptical galaxies and clusters \cite{BCAbell92,buoteCanizares,BC_NGC1332_96,BC_NGC3923_98, Buote02}.

Applying this method to the NGC 720 elliptical galaxy using Chandra X-ray data, Ref. \cite{Buote02} determined the ellipticity of the DM halo was roughly $\epsilon\approx  0.35-0.4$. Simulations for cross-sections of $\sigma/m = 0, 0.1, 1$ cm$^2$/g showed that the NGC 720 ellipticity was consistent with an interaction cross-section over mass of $\sigma/m\sim 0.1$ cm$^2$/g, as well as with CDM ($\sigma/m=0$ cm$^2$/g) \cite{Peter2013}. While these results presented convincing evidence that $\sigma/m=1$ cm$^2$/g was incompatible with the NGC 720 observation, strong assertions for lower cross-sections are difficult to make given the singular observation and significant scatter in the ellipticities of the simulated halos. Expanding on the NGC 720 results by performing a shape analysis of an ensemble of elliptical galaxies can potentially lead to more concrete statements. Still, the results of the X-ray ellipticity measurements of NGC 720 have been useful in applications to a wide range of DM models including hidden sector hydrogen \cite{atomicDM}, dissipative and ``double-disk'' dark matter \cite{fan2013}, charged DM \cite{mcdermott,feng09,MCDA}, as well as alternative gravity theories for DM phenomena such as MOND \cite{MONDellipticals}. 

In the present study, we expand on the scope of the analysis of Ref.~\cite{Buote02} by analyzing a sample of elliptical galaxies while additionally leveraging the capabilities of both the Chandra and XMM telescopes. Here we aim to build upon the previous X-ray shape measurements and provide data on the ellipticities of $M\sim 10^{12}-10^{13}M_{\odot}$ mass DM halos that will impact future studies of SIDM models especially in connection with future, higher-resolution N-body simulation of SIDM cosmologies.

The remainder of this paper is organized as follows. In section \ref{sec:theory} we describe the calculations needed to relate the mass distribution to the X-ray ellipticities, including discussion of the hydrostatic equilibrium condition. In section \ref{sec:sampleData} we present the criteria we utilized to create our galaxy sample selection and describe the Chandra and XMM data reduction procedures. We describe the analysis of the processed data including the ellipticities and surface brightness profile calculations and fitting procedures in section \ref{sec:analysis}. We present the results of these procedures in section \ref{sec:results} and discuss the implications of them for the DM self-interaction cross-section. Finally we present our conclusions in section \ref{sec:conclusion}.

%% Theory/Method
\section{\label{sec:theory}X-ray Emissivity as a Tracer of the Mass Distribution}

\subsection{Gravitational Potential of an Ellipsoidal Mass Distribution}\label{sec:potential}
For this study we use the ``X-ray Shape Theorem'' \cite{BinneyStrimpel,BCAbell92,buoteCanizares,BC_NGC1332_96,BC_NGC3923_98, Buote02} in order to determine the shape of the gravitating mass through observations of the X-ray emitting gas. This approach relies on the assumption that the gas is in hydrostatic equilibrium and therefore traces the gravitational potential. Thus, it is necessary to determine the gravitational potential produced by the total assumed mass distribution. The gravitational potential for an ellipsoidal mass distribution, $\rho(a)$, is given by the expression \cite{BT, BuoteHumphrey2012}

\begin{equation}
    \Phi(a) = -\pi G p q \int_0^{\infty}\frac{\psi\left(a^2(\tau,\v{x})\right) \:  d\tau}{\sqrt{(\tau + 1)(\tau + p^2)(\tau + q^2)}}
\end{equation}
where,
\begin{equation}
a(\tau, \v{x})^2 = \frac{x^2}{\tau+1}+ \frac{y^2}{\tau+p^2} + \frac{z^2}{\tau+q^2}
\end{equation}
and
\begin{equation}
\psi\left(a^2(\tau,\v{x})\right) = \int^{\infty}_{a^2(\tau,\v{x})} \rho(\tilde{a}^2)d\tilde{a}^{2}.
\end{equation}

\noindent
The elliptical radius $a$ is defined as 

\begin{equation}
a^2 = x^2 + \frac{y^2}{p^2} + \frac{z^2}{q^2}.
\end{equation}

\noindent
In this notation, the principal axes $a$, $b$, and  $c$ are assumed to be aligned along the $x$, $y$, and $z$ axes respectively with the values $p=b/a$ and $q=c/a$ being the axis ratios. The ellipticity is defined as $ \epsilon = 1-p = 1- b/a$. As we are interested in the flattening of the halo profile, we consider the two cases of oblate and prolate spheroids defined by $q=p$ (oblate) and $q=1$ (prolate) with the axis of symmetry along the line of sight. This in effect brackets the range of projected ellipticities of a triaxial ellipsoid. We take into consideration the following three mass distribution densities in our analysis: (i) a Navarro-Frenk-White (NFW) profile \cite{NFW97}, (ii) a Hernquist profile \cite{hernquist}, and (iii) a pseudo-isothermal profile (pIso). The functional form of the aforementioned profiles are explicitly given by

\begin{align}
    &\text{NFW: } &\rho \propto \frac{1}{a(a_s+a)^2} \label{eq:NFW}\\
    &\text{Hernquist: }&\rho\propto \frac{1}{a(a_s+a)^3}\label{eq:H}\\
    &\text{pIso: } &\rho \propto \frac{1}{(a_s^2+a^2)\label{eq:PL}}
\end{align}
where $a_s$ is the scale radius.

\subsection{\label{sec:HE} Hydrostatic Equilibrium --- Gas Density and X-ray Emissivity}
The isolated, relaxed elliptical galaxies are assumed to be in a state of hydrostatic equilibrium so that we may treat the X-ray gas as a tracer of the potential. This is expressed as a balance of the forces from internal gas pressure and gravitation given by the relation:

\begin{equation}\label{eq:isothermal}
\grad P_{gas} = -\rho_{gas} \grad \Phi,
\end{equation}
where $P_{gas}$ is the gas pressure, $\rho_{gas}$ is the gas mass density and $\Phi$ is the total gravitational potential. Taking the curl of both sides yields $(\grad \rho_{gas}) \times (\grad \Phi) = 0$, implying that surfaces of constant gas density are also surfaces of constant gravitational potential. For approximately isothermal gas distributions, the X-ray emissivity $(j_X)$ is related to the gas density as $j_X \propto \rho_{gas}^2$ \cite{buoteCanizares, BC_NGC1332_96}. Since surfaces of constant $\rho_{gas}$ are isopotential surfaces, it is also true that surfaces of constant $\rho_{gas}^2$, and consequently, surfaces of constant $j_X$ are isopotential surfaces as well \cite{BinneyStrimpel, BCAbell92,BC_NGC1332_96,BC_NGC3923_98, buoteCanizares, Buote02}. In practice, the observable quantity is not the 3-D emissivity itself, but rather the X-ray surface brightness, $\Sigma_X$, which is the 2-D projection of the emissivity along the line of sight and is given in terms of the gas density by the relation:

\begin{equation}
    \Sigma_X \propto \int_{los} j_X \propto \int_{los} \rho_{gas}^2.
\end{equation}
For an isothermal gas, Eq.~(\ref{eq:isothermal}) can be solved for the the gas density \cite{BT, buoteCanizares}:
\begin{equation}
    \rho_{gas}(a) = \rho_{gas,\:0}\exp\left[- \frac{\mu m |\Phi_0|}{k_b T}\left(1-\frac{\Phi(a)}{\Phi_0}\right)\right].
\end{equation}
We express the equation above more compactly as 
\begin{equation}
    \tilde{\rho}_{gas}(a) = \exp\left[- \Gamma\left(1-\tilde{\Phi}(a)\right)\right],
\end{equation}
where the tildes denote the dimensionless form for the expression normalized at the galaxy center and $\Gamma = \mu m |\Phi_0|/k_b T$. 

\begin{figure*}
\centering
    \includegraphics[width=0.32\linewidth]{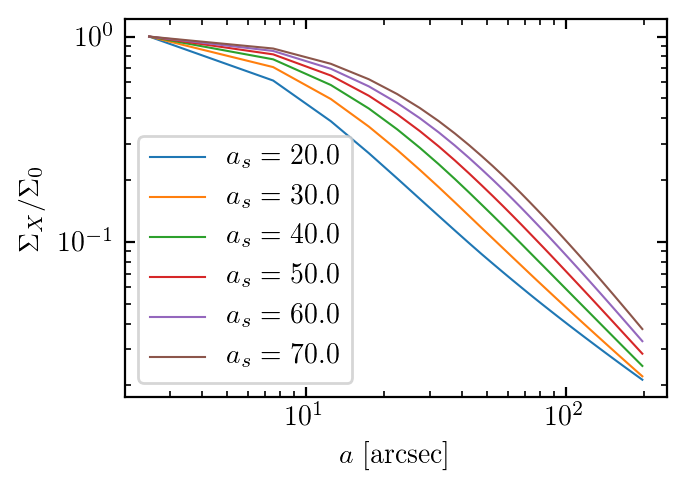}
    \includegraphics[width=0.32\linewidth]{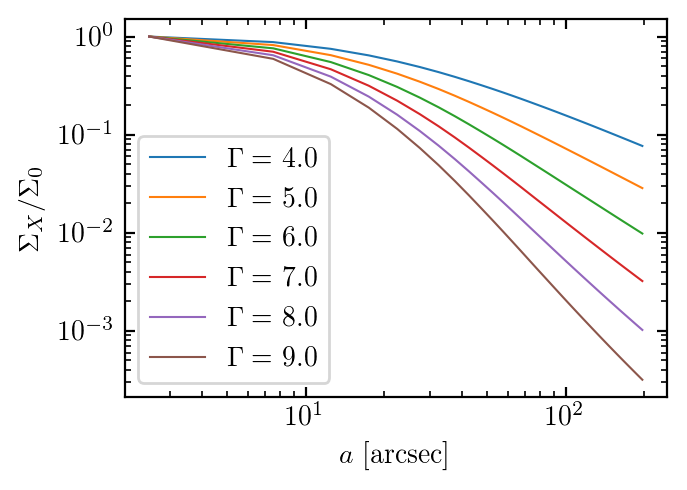}
    \includegraphics[width=0.32\linewidth]{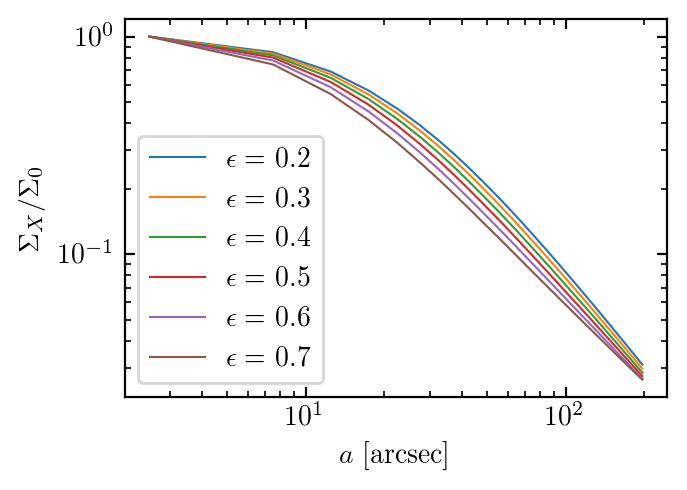}
\caption[Model X-ray Surface Brightness Profiles]{Normalized surface brightness profiles for a fiducial NFW dark matter density profile with $a_s=30^{''},\: \Gamma = 5$, and $\epsilon=0.4$ while varying $a_s$ {\it (left)}, $\Gamma$ {\it (middle)}, and $\epsilon$ {\it (right)}. }\label{fig:SBP_vary} 
\end{figure*}

\begin{figure*}
\centering
    \includegraphics[width=0.32\linewidth]{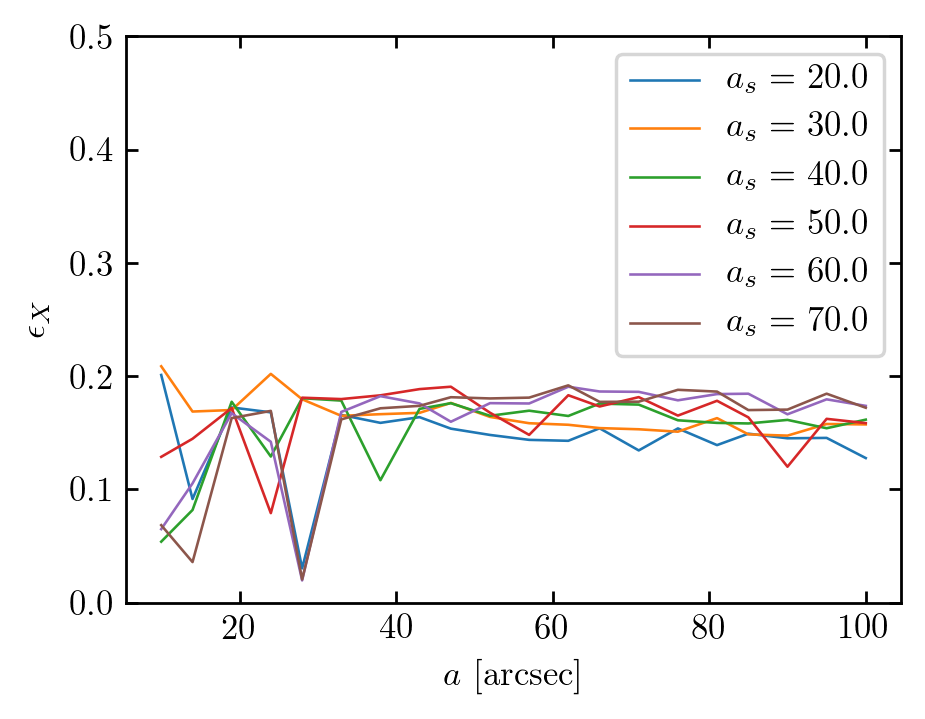}
    \includegraphics[width=0.32\linewidth]{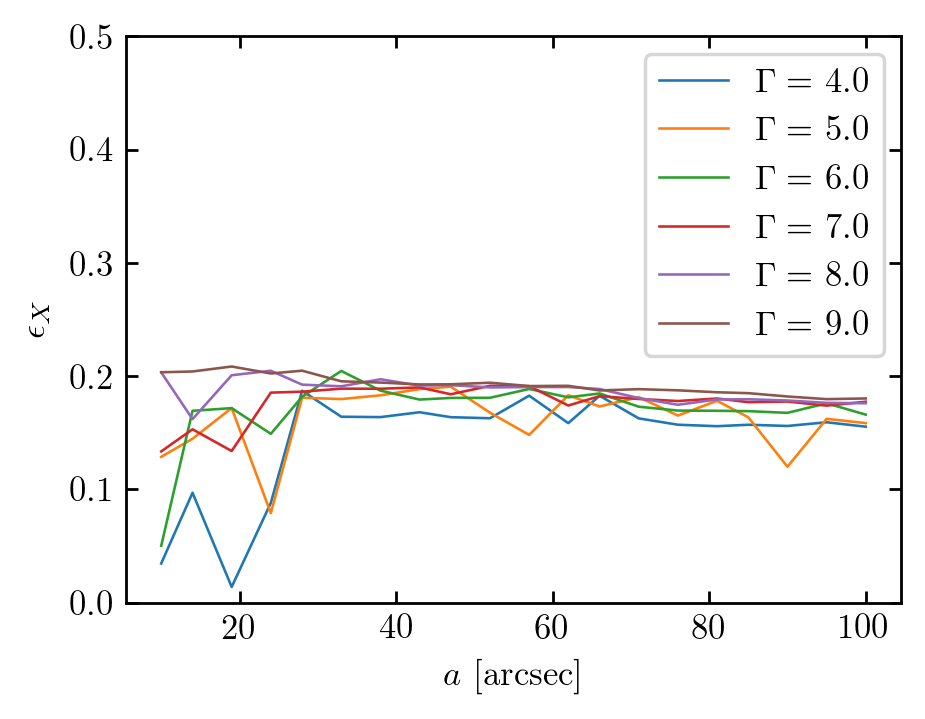}
    \includegraphics[width=0.32\linewidth]{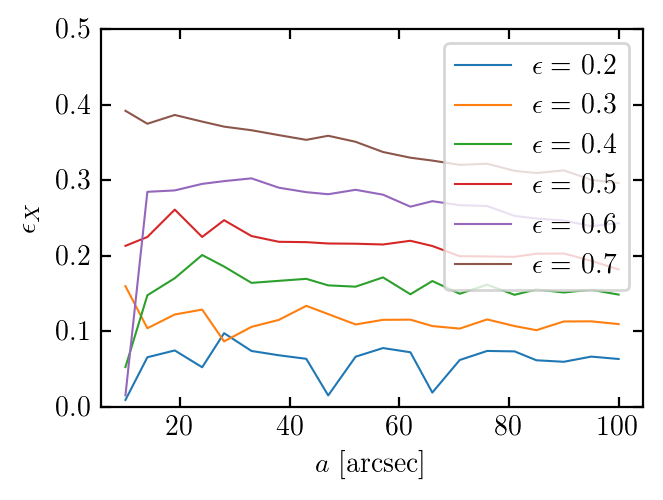}
\caption[Model X-ray Ellipticity Profiles]{Ellipticity profiles for a fiducial NFW galaxy with $a_s=30^{''},\: \Gamma = 5$, and $\epsilon=0.4$ while varying $a_s$ {\it (left)}, $\Gamma$ {\it (middle)}, and $\epsilon$ {\it (right)}. }\label{fig:ellip_vary} 
\end{figure*}

Once a model for the mass distribution (Eqs. \ref{eq:NFW}-\ref{eq:PL}) has been chosen, the parameters of interest in modeling the X-ray emission of a galaxy are $a_s,\: \Gamma$, and $\epsilon$, along with an appropriate normalization of the surface brightness. In each panel of figures \ref{fig:SBP_vary} and \ref{fig:ellip_vary} we show the the individual effects of varying each of these parameters on the surface brightness profiles and ellipticity profiles respectively. For illustrative purposes only, we adopt base parameter values of $a_s=30^{''},\: \Gamma = 5$, and $\epsilon=0.4$ and assume an NFW profile. While the surface brightness profile unsurprisingly has a strong dependence on $a_s$ and $\Gamma$, the effects of varying the ellipticity of the halo are less pronounced. In the X-ray ellipticity profile, the $a_s$ and $\Gamma$ parameters have almost no discernible effects aside from a slight decrease in observed $\epsilon_X$ at the lowest values ($a_s\sim 20$'', $\Gamma \sim 4$). Naturally, the ellipticity of the matter distribution used in the model has a major impact on the X-ray ellipticity profile. In addition, for more elliptical mass distributions the X-ray profile also tends to exhibit a radial dependence.

%%%%%%%%%%%%%%%%%%%%%%%%%%%%%%% Modeled SB profiles %%%%%%%%%%%%%%%%%%%%%%%%%%%%%%%

%%%%%%%%%%%%%%%%%%%%%%%%%%%%%%% Modeled ellipticity profiles %%%%%%%%%%%%%%%%%%%%%%%%%%%%%%%

%%%%%%%%%%%%%%%%%%%%%%%%%%%%%%% Modeled SB profiles %%%%%%%%%%%%%%%%%%%%%%%%%%%%%%%
\section{Galaxy Sample and Data Reduction} \label{sec:sampleData}
%%%%%%%%%%%%%%%%%%%%%%%%%%%%%%% Data Sample %%%%%%%%%%%%%%%%%%%%%%%%%%%%%%%
\subsection{Galaxy Selection}
The selection of galaxies used in this analysis is guided by a few conditions that justify the assumption of hydrostatic equilibrium. We focus our attention on relaxed, isolated elliptical galaxies. Thus, we seek candidates that exhibit regular, circular, or elliptical X-ray morphologies as expected for a suitably relaxed galaxy \cite{chandra_galaxy_selection, BH_review}. In particular, we avoid galaxies with bright central AGN or significant AGN feedback. The elliptical galaxy NGC 4374 (M84) \cite{M84} provides an example case where this relaxation condition does not hold due to significant disruption of the X-ray gas from AGN feedback. The isolation criteria include 

(i) no close neighbors, 

(ii) no signs of interaction, and 

(iii) no evidence for a recent merger.  

\noindent Additionally, since the stellar mass dominates the overall mass distribution in the central regions, we seek galaxies for which existing observational data extends far beyond the effective stellar radius. This is necessary in order to probe the underlying mass distribution in the outer regions where the dark matter component is the dominant contribution to the gravitational potential. The galaxies that fit these criteria and are used in this analysis are listed in Table~\ref{tab:gal_sample} along with some basic properties including their distance and K-Band effective radius taken from the Two Micron All Sky Survey (2MASS) Extended Source Catalog (XSC) \cite{2mass}.
\begin{table*}[t]
\centering
\def\arraystretch{1.25}
\setlength{\tabcolsep}{12pt}

\begin{tabular}{lccc@{\hskip 18 pt}c@{\hskip 18 pt}cc}
	\hline\hline
	Galaxy & Dist. (Mpc)&  ''/kpc&  $r_e$ (kpc)  \\

	\hline
   
IC4451 & 55.8 \cite{springbob}& 0.27 & 4.07 \\
IC4956 & 70.1  \cite{tf_dist}& 0.34 & 4.59 \\
NGC1521 & 69.5 \cite{tully}& 0.34 & 5.53 \\
NGC4125 & 24.0 \cite{tully}& 0.12 & 3.65 \\
NGC4555 & 110 \cite{saulder}& 0.53 &  5.97 \\
NGC57 & 70.1 \cite{tf_dist}& 0.34 & 4.98 \\
NGC6482 & 59.2 \cite{ma2014}& 0.29 &  3.47 \\
NGC7785 & 48.4 \cite{tf_dist}& 0.23 & 4.05 \\
NGC7796 & 51.3 \cite{tully}& 0.25 & 4.03 \\
NGC953 & 66.3 \cite{crook}& 0.32 &  2.91 \\
NGC720 & 25.7 \cite{tonry} & 0.12 &  3.14 \\
\hline\hline
\end{tabular}
\caption{Sample of galaxies used in this analysis. References for the distances are provided. The effective radius $r_e$ is the K-band effective radius taken from the 2MASS XSC catalog \cite{2mass}. \label{tab:gal_sample} }
\end{table*}

\newpage
\begin{table*}[tbph]
\centering
\def\arraystretch{1.5}
\setlength{\tabcolsep}{12pt}

\begin{tabular}{lccc@{\hskip 18 pt}c@{\hskip 18 pt}cc}
	\hline\hline
	&\multicolumn{2}{c}{XMM}&\multicolumn{2}{c}{Chandra}\\
	\cline{2-5}
	&MOS1 + MOS2&&\\
	Galaxy&ObsID &Exp$_{1+2}$ (ks)& ObsID  & Exp. (ks) \\

	\hline
\multirow{2}{*}{IC4451}&
0503480501 &4.26 + 6.02 & 13808 & 25.05 \\
&  0673080101 &  66.04 + 66.58&   --  & --\\
\hline
\multirow{2}{*}{IC4956} & 
 0503480801 &  0.96 + 3.38& -- &-- \\
	& 0693190401 & 81.44 + 86.74& -- &-- \\

\hline
NGC1521                  & 0552510101 & 98.15 + 100.82& 10539 & 40.79 \\
\hline
NGC4125 			 & 0141570201 & 98.15 + 100.82 & 2071 & 54.76 \\
\hline
NGC4555 				 & 0403100101 &62.64 + 62.26 & 1884 & 24.44 \\
\hline
NGC57 					 & 0202190201 & 21.17 + 21.23 & 10547 & 8.69 \\
\hline
\multirow{8}{*}{NGC6482}&
0304160401 & 7.92 + 7.9 & 3218  & 18.94 \\
						   & 0304160801 &  5.21 + 5.39 & 19584 & 23.50 \\
						   &     --     & -- & 19585 & 17.00 \\
						   &     --     & -- & 20850 & 17.97 \\
						   &     --     & -- & 20857 & 20.00 \\
						   &     --     & -- & 20978 & 17.97 \\	
						   &     --     & -- & 20979 & 8.44  \\
						   &     --     & -- & 20980 & 67.77 \\
\hline
NGC7785 				  & 0206060101 & 16.5 + 17.08 & -- & --\\
\hline
\multirow{2}{*}{NGC7796}&
0693190101 & 47.64 + 56.74 & 7401 & 16.91 \\
						   &     --     &     --         & 7061 & 44.20 \\
\hline
\multirow{2}{*}{NGC953}&
0722360201 & 69.49 + 75.84 & 11262 & 5.92 \\
						   & 0762220101 & 95.55 + 95.04 & 14899 & 37.03 \\
\hline
NGC720 					& 602010101 & 81.76 + 81.71 & 492 & 21.69 \\

\hline\hline
\end{tabular}
\caption[X-ray Observations of Elliptical Galaxies]{Sample of galaxies and their corresponding ObsIDs and cleaned exposure times from XMM and Chandra (see section \ref{sec:data_reduct} for details).  For the XMM exposure time, we show the individual cleaned exposure for each of the MOS1 and MOS2 EPIC cameras. \label{tab:gal_exp} }
\end{table*}
\newpage

%%%%%%%%%%%%%%%%%%%%%%%%%%%%%%% Data reduction %%%%%%%%%%%%%%%%%%%%%%%%%%%%%%%
\subsection{Observations and Data Reduction}\label{sec:data_reduct}
%%%%%%%%%%%%%%%%%%%%%%%%%%%%%%% Data reduction - XMM %%%%%%%%%%%%%%%%%%%%%%%%%%%%%%%
\subsubsection{XMM-Newton}
Each galaxy in our sample has at least one observation with the XMM-Newton telescope. We use archival data from the EPIC MOS1 and MOS2 cameras in the soft X-ray energy band $0.5-2$ keV. The processing of the data is performed with the Science Analysis Software (SAS\footnote{\href{http://xmm-tools.cosmos.esa.int/external/xmm_user_support/documentation/sas_usg/USG/SASUSG.html}{"Users Guide to the XMM-Newton Science Analysis System", Issue 15.0, 2019 (ESA: XMM-Newton SOC).}} v16.1.0) and the Extended Source Analysis Software (ESAS) \cite{ESAS_AAS} following the steps outlined in the ESAS Cookbook\footnote{\url{https://heasarc.gsfc.nasa.gov/docs/xmm/esas/cookbook/xmm-esas.html}} for diffuse emission. Specifically, we run the \texttt{emchain} program to prepare the events list products for use with the ESAS tasks. The \texttt{mos-filter} routine (which in turn calls the SAS routine \texttt{espfilt}) is used to remove periods of high background and determine the good time intervals (GTI). Light curves for each observation and MOS camera were inspected manually to ensure quality and observations that were not adequate (see e.g. the examples in the ESAS cookbook) are not included in this work. Nearly all observations exhibited some periods of high background that were subsequently removed. The remaining clean exposure times can be found in table \ref{tab:gal_exp}. Images and exposure maps are then created for each observation and MOS camera using the \texttt{mos-spectra} routine and then combined. During the ESAS processing, images are binned to pixel sizes of $\sim 2.5 '' \times 2.5 ''$.

\subsubsection{Chandra}
For nine out of the 11 galaxies in our sample there exists archival Chandra data that is suitable for this analysis. We restrict our analysis to the ACIS S3 chip and again consider the $0.5-2$ keV energy band. The data are processed with the CIAO 4.11\footnote{\url{https://cxc.cfa.harvard.edu/ciao/}} \cite{CIAO} software along with the corresponding Chandra calibration database (CALDB) v4.8.2 following the standard procedure for Chandra data. The Chandra data are binned into pixels of size $\sim 1$''$\times 1$''. For flare removal we use the \texttt{lc\_clean} routine with default parameters and again manually inspect the light curves before proceeding. The remaining cleaned exposure times are shown for each ObsID in \ref{tab:gal_exp}.

\subsubsection{Point Source Removal}
\begin{figure*}[!htbp]
\centering
\includegraphics[width=0.8\textwidth, height=0.9\textheight]{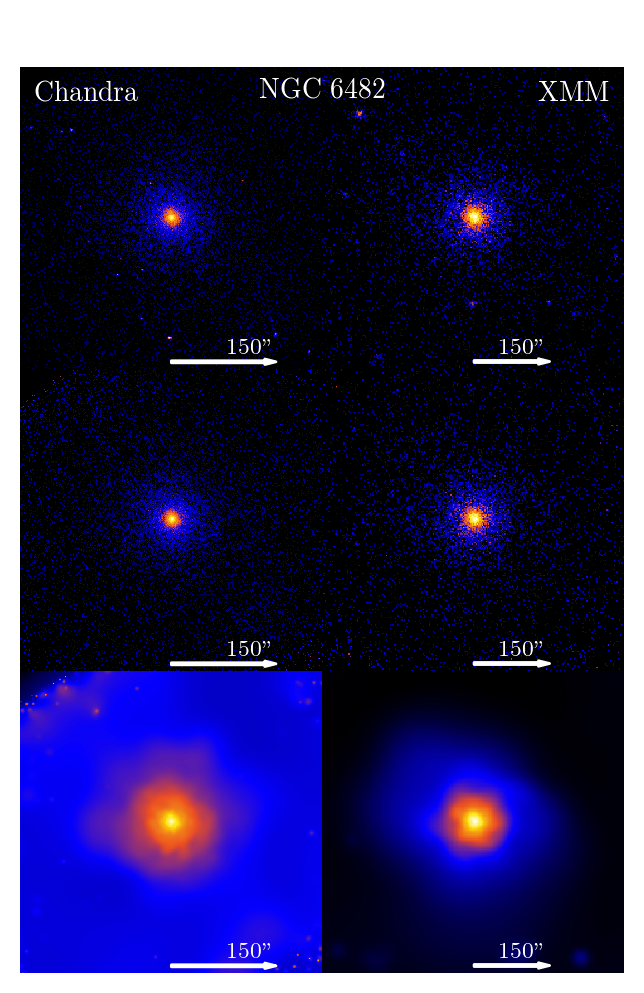}
\hfill
\caption[NGC 6482 Observations from Chandra and XMM]{Chandra (left column) and XMM-Newton (right column) images of NGC6482. The top row is prior to point source detection. The middle row is after running the \texttt{wavdetect} and \texttt{dmfilth} routines to remove points sources. Finally, the bottom row shows the point source cleaned, smoothed, and exposure corrected images.}
\label{fig:proc_data}
\end{figure*}

Since we are interested in the extended diffuse emission of the galaxies, it is necessary to remove bright point sources. Further, simply removing or masking the point sources is insufficient as the empty source regions in the image can significantly affect the ellipticity measurements. We therefore need to both identify the point sources and reasonably model the true diffuse emission in their place. For the identification of the point sources we use the \texttt{wavdetect} CIAO routine which provides a wavelet function source detection method. We supply this routine with the point spread function (PSF) map built by running \texttt{fluximage} (and thus, \texttt{mkpsfmap}) for the Chandra data. In the XMM data we use a psfmap with a constant size of 5''. The exposure maps created by \texttt{fluximage} and \texttt{mos-spectra} are used for the Chandra and XMM images, respectively. Identified source regions were then removed and filled using the CIAO \texttt{dmfilth} task which takes as input source and background regions produced by the \texttt{roi} routine. When running \texttt{dmfilth} we use the \texttt{POISSON} method, which replaces the source region by sampling from the Poisson distribution whose mean is that of the pixel values in the background region. In figure \ref{fig:proc_data} we show example Chandra and XMM images of NGC 6482 before source detection, after source detection and replacement, as well as an image smoothed using the CIAO task \texttt{csmooth}.

%%%%%%%%%%%%%%%%%%%%%%%%%%%%%%% Data Images %%%%%%%%%%%%%%%%%%%%%%%%%%%%%%%
%Might be good to make this a 4 panel or something with PS, PS removed, then smooth for the two instruments

\section{X-ray Ellipticity and Brightness Profiles}\label{sec:analysis}
The ellipticities of the X-ray images for each galaxy in our data set are calculated using the image moments method as outlined in \cite{trumpler, McMillan, buoteCanizares, Buote02, carterMetcalfe} which we briefly review here. We begin by finding the centroid of a circular aperture at the desired semi-major radius $a$. The centroid of the radius is then calculated from the first order moment and the aperture shifted to that point. This process is repeated until the centroid shift changes less than some tolerance (roughly $\sim 1$ pixel). With the centroid found, we again start with a circular aperture of radius $a$, and calculate the second order image moments. These are effectively the elements of the inertia tensor of the image within the aperture, and allow the determination of the ellipticity within the aperture and orientation angle of the semi-major axis \cite{trumpler, McMillan, buoteCanizares,  carterMetcalfe}. This process is performed iteratively until the measured ellipticities and orientations converge. For the error estimation of the ellipticity profile, we follow the procedure of \cite{Buote02} and use a Monte Carlo approach. The pixel counts are assumed to follow Poisson statistics, so we create a simulated image by sampling the pixel values from a Poisson distribution with the original pixel count as the mean. We then calculate the ellipticity for each radii $a$ as in the case of the original data image, repeating this for 100 instances. The standard deviation of the samples generated through this process is taken to be the $1\sigma$ error of our ellipticity.

In addition to the ellipticity radial profile, we also require the radial surface brightness profile in order to characterize the shape of the gravitating mass distribution. This is calculated by adding the counts within several annuli and dividing by the area of each annulus. Errors are calculated assuming Gaussian statistics (i.e. $\sigma_i \sim \sqrt{N_i}$ for $N_i$ counts in annulus $i$) based on the lowest counts per bin in our samples being on the order of $\gtrsim 100$. 

To perform the fitting procedure we begin by generating a model image for each assumed mass distribution and either prolate or oblate configuration based on the calculations in sections \ref{sec:potential} and \ref{sec:HE}. The ellipticity and surface brightness profiles are calculated for the model image and a $\chi^2$ statistic is used to determine the fit to the data. We then minimize the $\chi^2$ statistic with $a_s, \: \Gamma,$ and $\epsilon$ as free parameters using a Nelder-Mead simplex algorithm \cite{NelderMead, numericalRecipes} with a dimension dependent implementation of the expansion, contraction, and shrink parameters \cite{Gao2010}.

\begin{figure}
\centering
\includegraphics[width=0.75\textwidth]{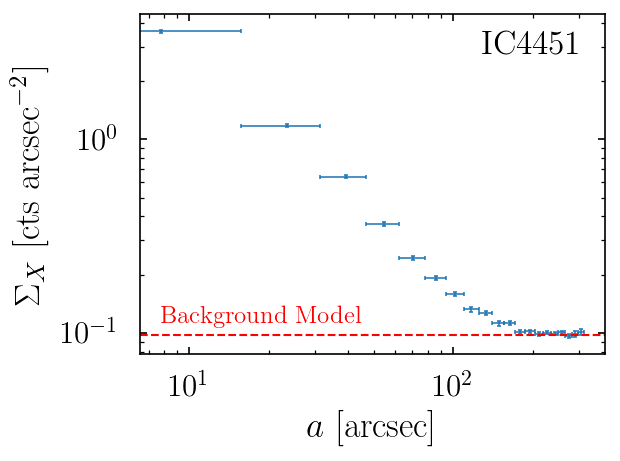}
\hfill
\caption[Surface Brightness Profile and Background Model of IC 4451]{Surface brightness profile of IC 4451 without background subtraction. The red line shows the background model used in later portions of the analysis. }
\label{fig:bg_model}
\end{figure}

In the process described above, we treat the background modeling in a slightly different manner for the ellipticity and brightness profiles. The constant background model is estimated from the flattening of the brightness profile at large radii. This is illustrated in figure \ref{fig:bg_model}. For the ellipticity measurements, we subtract this background from the data and fit to the model generated profiles.

We note that problems can arise in the larger radius bins if the profile is calculated from the background subtracted image due to over subtraction. Essentially, some annular bins may have average counts per pixel slightly less than the background model. To avoid these problems we model the background simultaneously with the calculated model image rather than subtracting from the data. For the ellipticity profiles, over-subtracting the observed counts is not an issue due to the elliptical aperture that contains the bright central regions and ensures that the average counts per pixel are greater than the background model.

\section{Results}\label{sec:results}
\begin{figure*}
\centering
\includegraphics[width=0.99\textwidth]{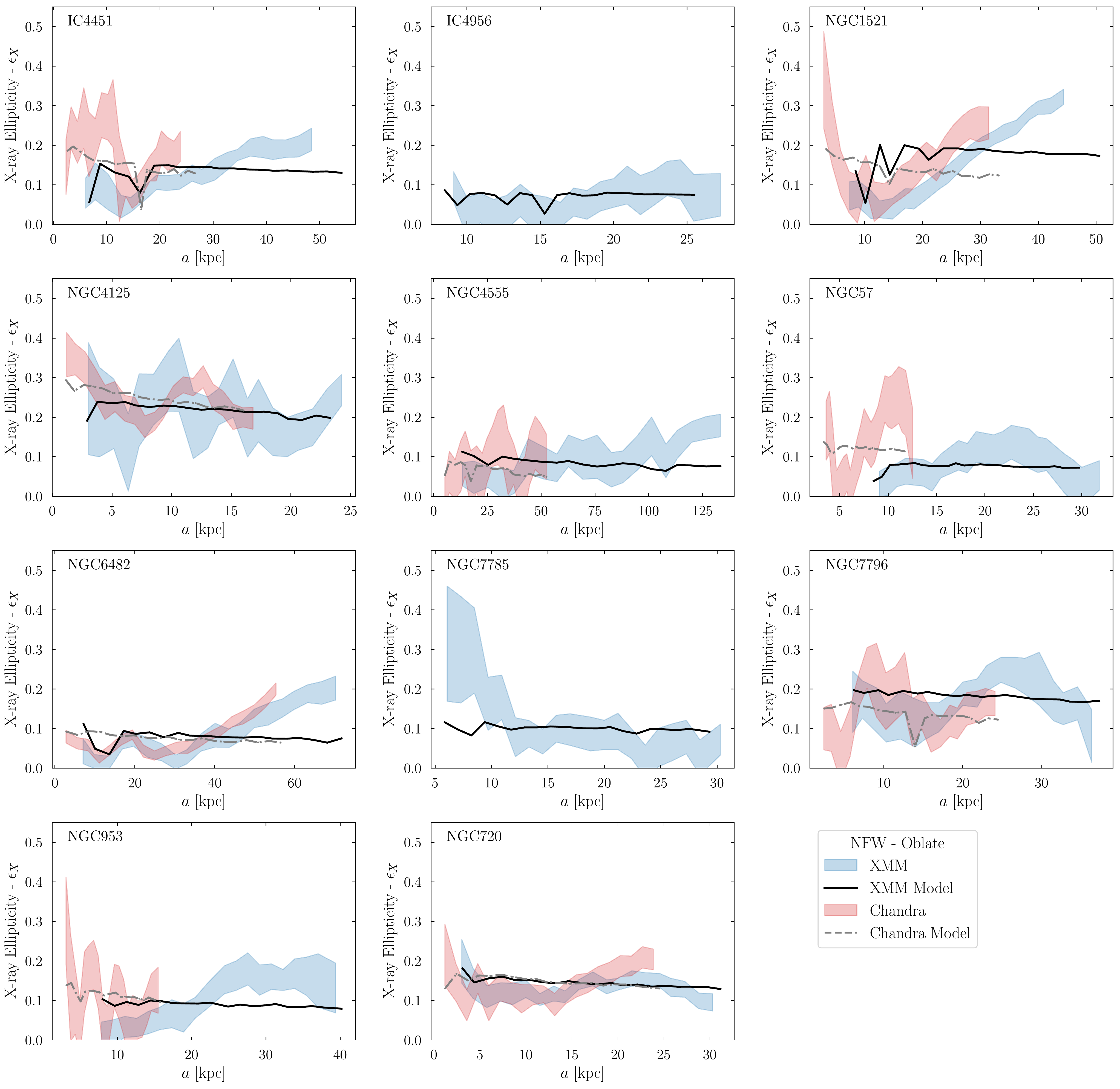}
\caption[Ellipticity Profiles of the Best-Fit NFW Models]{Ellipticity profiles of the best-fit models for each galaxy assuming an oblate NFW mass distribution plotted along with the profiles from the observational data.}
\label{fig:ellipData_NFW}
\end{figure*}

\begin{figure*}
\centering
\includegraphics[width=0.99\textwidth]{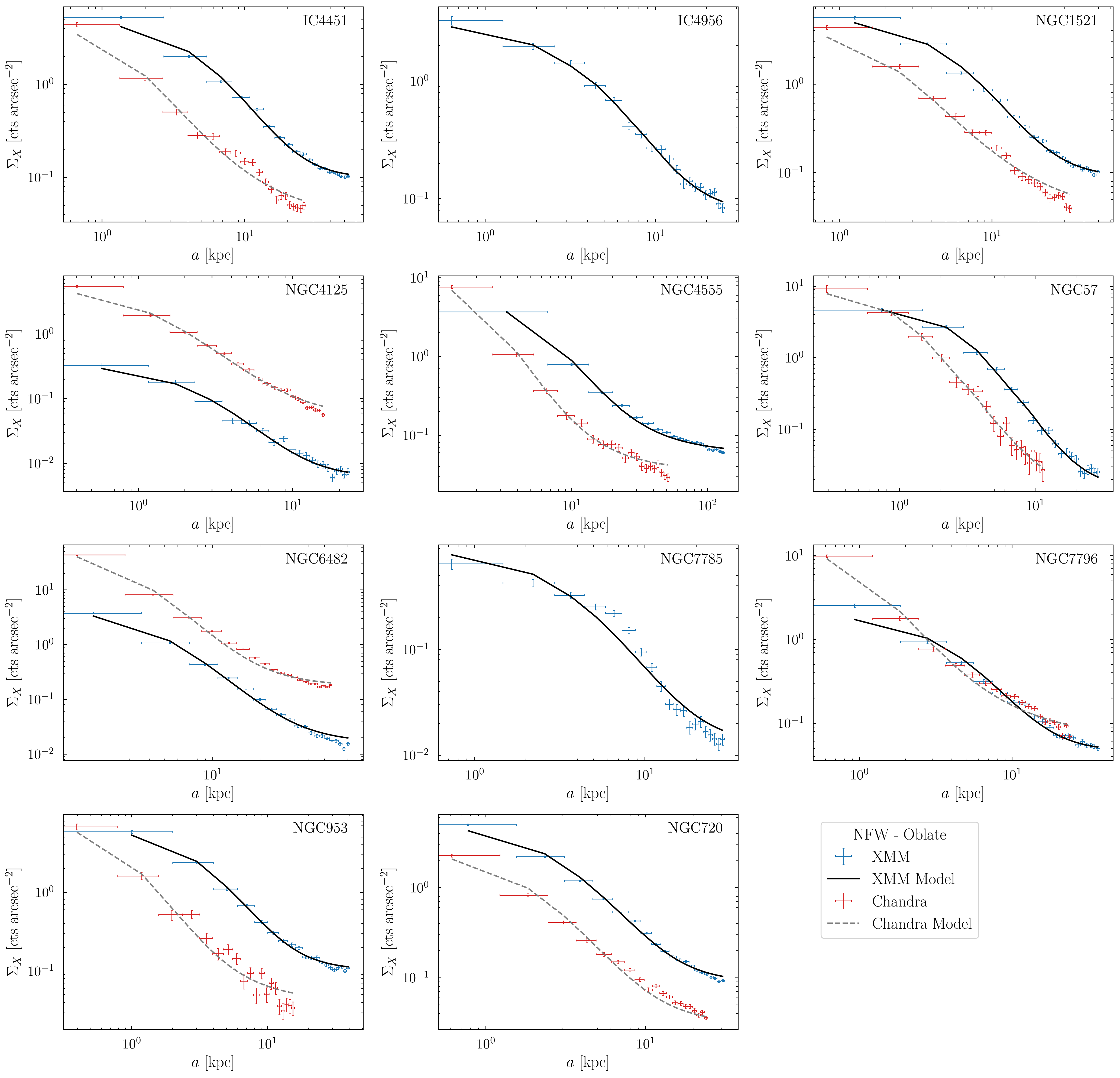}
\caption[Surface Brightness Profiles of the Best-Fit NFW Models]{Surface brightness profiles of the best-fit models for each galaxy assuming an oblate NFW mass distribution plotted along with the profiles from the observational data.}
\label{fig:SBPData_NFW}
\end{figure*}

\begin{figure*}
\centering
\includegraphics[width=0.45\textwidth]{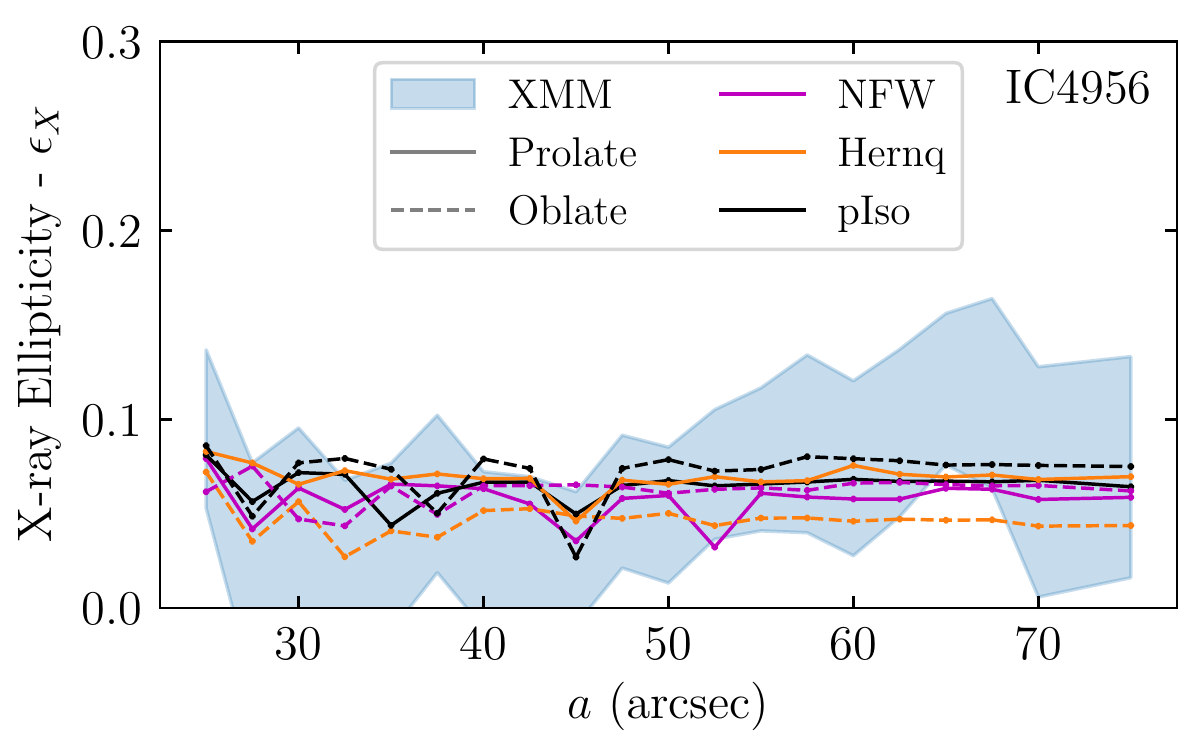}
\includegraphics[width=0.45\textwidth]{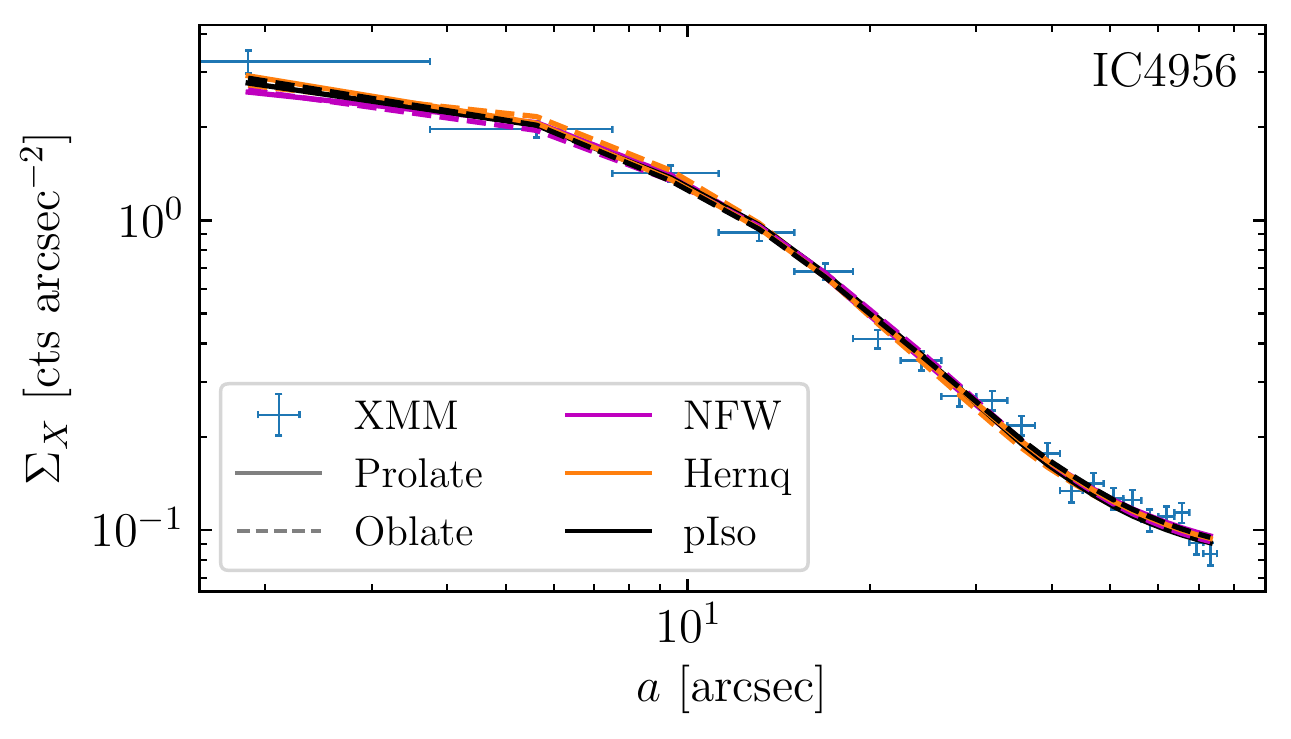}
\hfill
\caption[Ellipticty and Surface Brightness Profiles of IC 4956]{Best-fit ellipticity and surface brightness profiles for the XMM observations of IC4956. Here we show the best fitting profiles for each mass distribution configuration. Colors correspond to different profiles, while the solid and dashed lines refer to the prolate and oblate configurations respectively.}
\label{fig:IC4956}
\end{figure*}

In figures \ref{fig:ellipData_NFW} and \ref{fig:SBPData_NFW} we show the best fit profiles for an NFW oblate mass configuration overlaid on the data for each galaxy. Figure \ref{fig:IC4956} shows the best-fit profiles for each configuration overlaid on the IC 4956 XMM data. The best-fit results for the XMM and Chandra data are summarized in tables \ref{tab:xmm_fits} and \ref{tab:chandra_fits} respectively, along with the reduced $\chi^2$ values. In most cases, the $\chi^2$ statistic does not provide a quantitatively ``good'' fit to the data. In the case of the surface brightness profile this can at least partially be attributed to small error bars that amplify the $\chi^2$ value despite what appears to be a fairly strong qualitative agreement between the data and the model. This is similar to what was found in the study of NGC 720 from Ref. \cite{Buote02}. However, it is clear from looking at figure \ref{fig:ellipData_NFW} that a large contributor to the $\chi^2$ is the fact that the data for several of the galaxies do not follow the relatively flat ellipticity profiles of the model. This is not entirely surprising, since in this analysis we are assuming either prolate or oblate spheroidal mass-distributions with a constant ellipticity. More realistically, many of these galaxies are likely to be better represented by a triaxial distribution. They would therefore be expected to exhibit an isophotal ``twist'' and variations in their X-ray ellipticity profile \cite{buoteTwist, romanowsky}. Rotating cooling flows may also alter the X-ray isophotes and contribute to variations in the radial profiles \cite{buoteTwist}. Nevertheless, since we have used the prolate and oblate configurations as a way to bracket the triaxial case we treat the results as relatively representative of the underlying mass distribution, though keeping in mind the potential need for more sophisticated modeling of the halo shape and possible astrophysical activity.

Another point of note is the radial range chosen for the fitting procedure. In our basic approach the minimum of the range was chosen to be $10$ pixels for stability of the iterative moment method when calculating the ellipticity. The maximum radial limit was taken to be the point at which the emission region became background dominated (see figure \ref{fig:bg_model}) or the chip edge was reached. However, there are good reasons to restrict this range from both ends on a galaxy-by-galaxy basis. In the inner regions, the stellar component contributes significantly to the overall mass distribution. For some of the galaxies in our sample, previous studies have been able to model the mass distribution to give a detailed description of how the mass profile and its constituent components change with radius (for example, see \cite{7785_mass,6482_mass,720_mass,elixr_1521,ngc4555_mass,chandra_galaxy_selection}). In cases for which the full mass modeling has not been performed, it can be roughly assumed that the stellar mass has a non-negligible contribution within the effective radius $r_e$ \cite{brighenti,gerhard}. In choosing a more restrictive minimum radius these characteristics of the galaxies should be kept in mind.

At larger radii, it is less clear that there is a natural maximum limit smaller than the background limit. In the NGC 720 study from Ref. \cite{Buote02}, the upper radial limit $\sim 150''$ was chosen because of strange behavior observed at larger radii wherein the profile diverged from the values at smaller radii. Applying this to our sample there are potentially six galaxies (IC 4451, NGC 1521, NGC 57, NGC 6482, NGC 953, NGC 720) that exhibit this behavior to some extent for at least one of the observations. The origin of this divergence is unclear, although one physically motivated possibility is that the assumption of hydrostatic equilibrium does not accurately apply at these radii. 

However, Ref. \cite{Buote02} also points out that for their observation the region $\gtrsim 150''$ is near the CCD edge which may be causing the strange behavior, and that observations with a wider field of view such as XMM could give insight into whether this is an observational or physical effect. In some of our galaxies, we find that in these large radius regimes the Chandra data exhibits this behavior to a much greater degree and diverges from the XMM profile. For example, the Chandra ellipticity diverges from the XMM values for NGC 6482, NGC 1521, and NGC 720 at $a \gtrsim 40,20,18$ kpc respectively (note that for NGC 720, 18 kpc $\approx 150''$). This suggests the strange behavior at large radii is likely due to observational effects rather than a failure of hydrostatic equilibrium, and furthermore that removing regions at large radii exhibiting this behavior could improve the accuracy of the fits.

These considerations along with the ability to compare our results with other studies motivate us to also explore a more physically motivated range over which to perform the fitting procedure. We select the range $a = r_e \rightarrow 5r_e$, which roughly corresponds with the range used in both the NGC 720 analysis \cite{Buote02} as well as simulation results \cite{Peter2013}. In some cases data is not available at this range either due to the presence of the CCD edge or being background limited. In such cases we only fit to the extent of available data. The results of fitting in this range for the XMM and Chandra datasets are shown in tables \ref{tab:xmm_fits_range} and \ref{tab:chandra_fits_range} respectively. Broadly speaking, the results over this range tend to perform similarly, both in terms of fit parameters and $\chi^2_{red}$ though this varies on a galaxy-to-galaxy basis. While the $a_s$ and $\Gamma$ are quite consistent between the two fitting ranges, the ellipticities tend toward slightly lower values in the range-restricted setup. This may be attributable to avoiding the potentially anomalous behavior at large radii discussed above; however, it is difficult to draw strong conclusions given the smaller statistics of the limited fit range. A direct comparison between the two choices of radial range can be seen in the right panel of figure \ref{fig:ellip_vary2}. since the results of the two choices are not highly disparate and the background limited setup provides somewhat better fits, the remainder of this section will focus on these results. However, see section \ref{sec:sidm} for more discussion comparing the two choices.

\begin{figure*}
\centering
\includegraphics[width=0.75\textwidth]{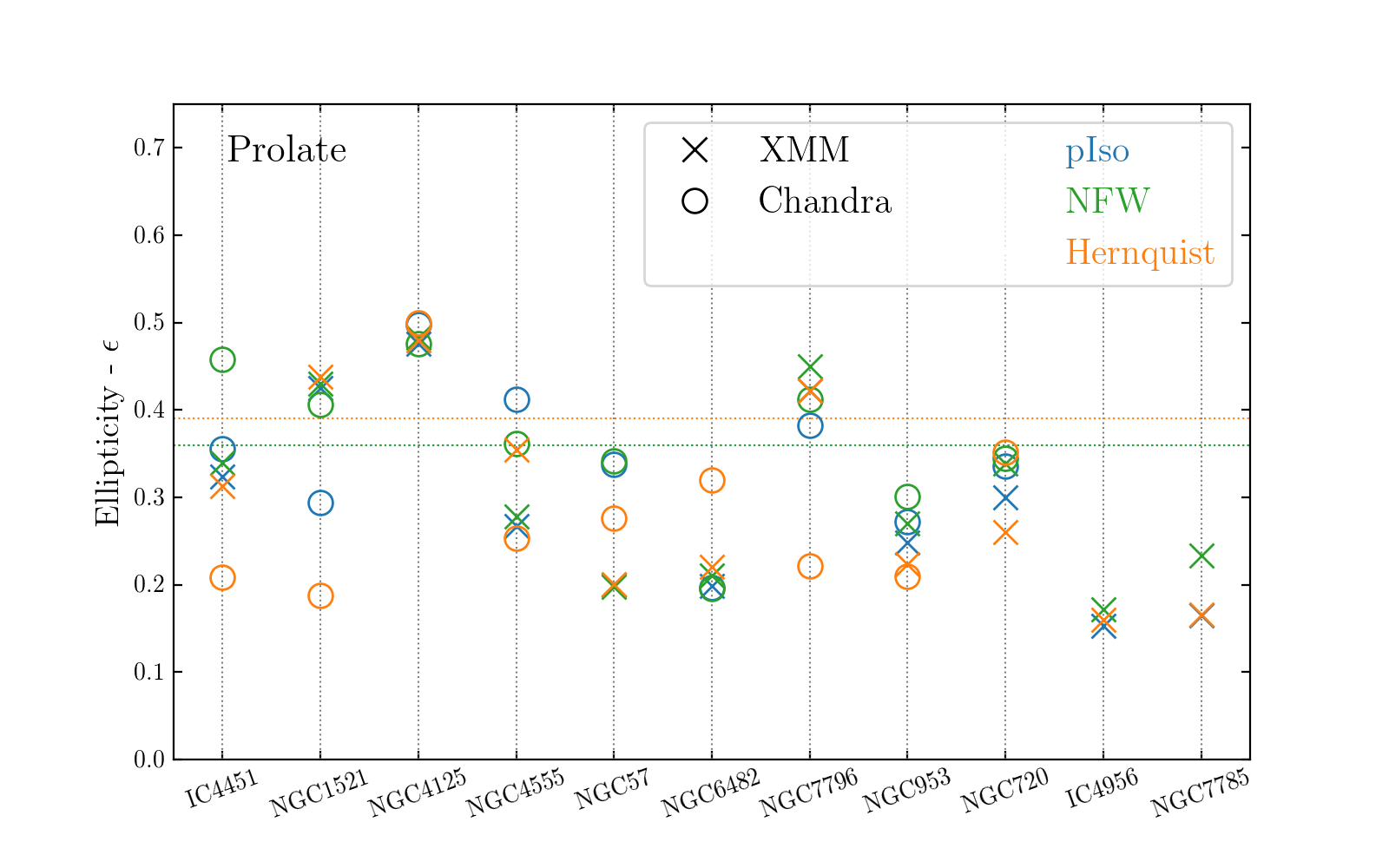}\\
\includegraphics[width=0.75\textwidth]{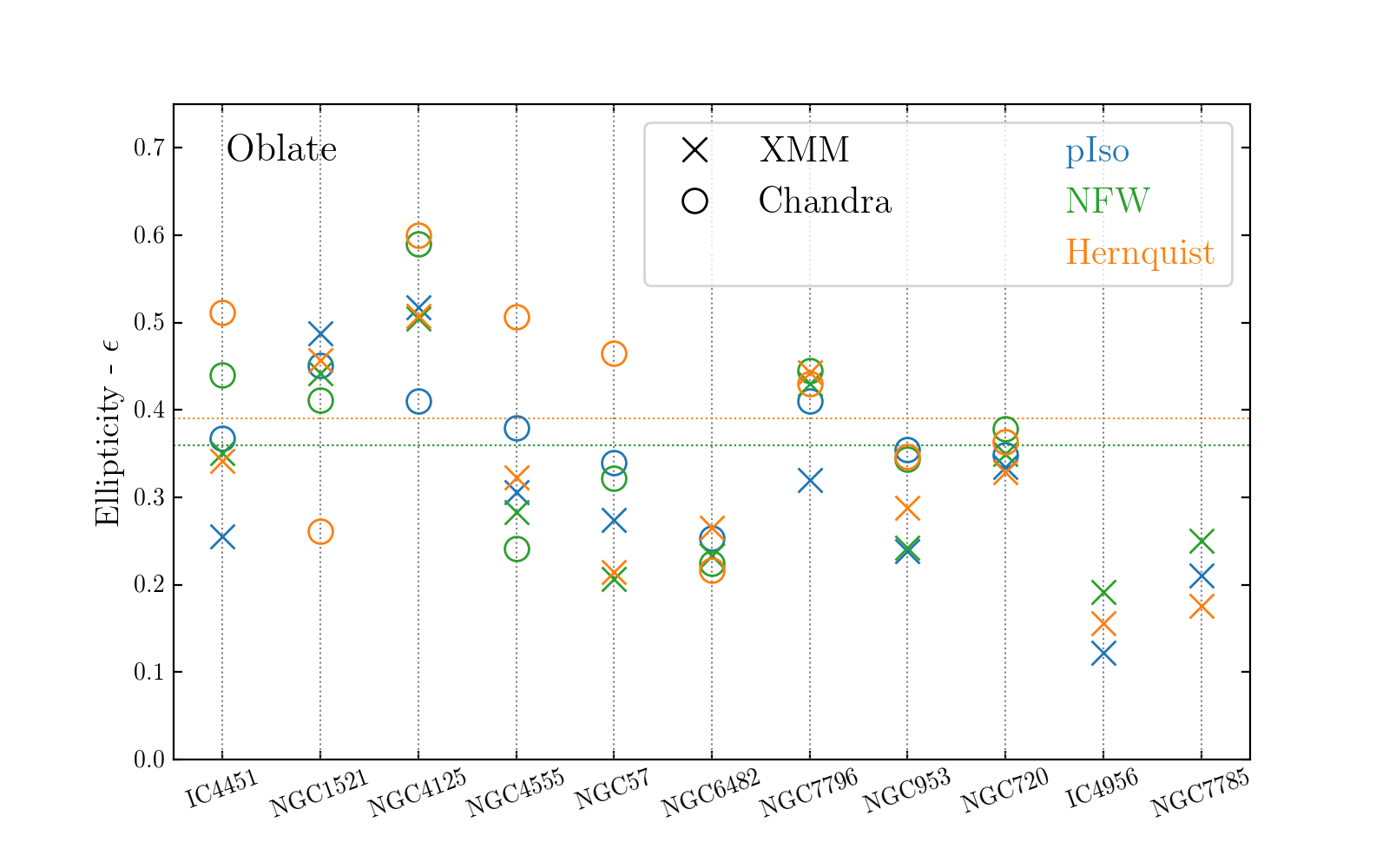}
\hfill
\caption[Ellipticities for Each Galaxy and Mass Configurations]{Best-fit DM halo ellipticities $\epsilon$ for each galaxy in our sample determined from the XMM (X's) and Chandra (circles) observations in the background limited range. The top panel shows the prolate configuration, while in the bottom panel we show the oblate configuration. The horizontal lines show the values found for NGC720 in \cite{Buote02} (Note that in the prolate configuration the ellipticities for the $\rho\sim a^{-2}$ and NFW halos are overlapping.) }
\label{fig:galaxy_ellips}
\end{figure*}

Fits to the data were fairly consistent across the various mass configurations for a given galaxy. The slope parameter $\Gamma$ varies from galaxy to galaxy ranging from $\sim 6-9$ but is otherwise roughly consistent across mass configurations. An interesting note is that the scale radius $a_s$ in the pIso models is almost always considerably smaller than for the NFW or Hernquist profiles. As the scale radius characterizes the core in the pIso profile, this seems to suggest that the X-ray emission prefers a small-core, nearly isothermal $\rho \propto a^{-2}$ profile. In figure \ref{fig:galaxy_ellips} we show the ellipticities for each galaxy and mass distribution configuration. For comparison, we also show the ellipticity for NGC 720 determined in \cite{Buote02} as horizontal colored lines. In general, their measured ellipticity falls comfortably within the range of the measured ellipticities of our ensemble. The values found in our analysis for NGC 720 tend to be slightly lower, although this could potentially be due to exposure corrections (see section 2 of \cite{Buote02}) or calibration effects. In addition, there does not appear to be any significant relationship between the ellipticity and the mass configuration for a given galaxy (e.g. NFW halos do not consistently result in the lowest ellipticities) .

%%%%%%%%%%%%%%%%%%%%%%%%%%%%%%% Profiles %%%%%%%%%%%%%%%%%%%%%%%%%%%%%%%

\subsection{Implications for Self-Interacting Dark Matter }\label{sec:sidm}
In line with early analytic arguments concerning SIDM \cite{SS}, simulations have demonstrated that the DM self-interactions produce halos with greater sphericity than in CDM \cite{Peter2013, DSS2000}. In Ref. \cite{Peter2013}, DM halos were simulated for cross-sections of $\sigma/m=$ 0 cm$^2$/g (CDM), 0.1 cm$^2$/g, and 1 cm$^2$/g. These simulations were originally motivated by the results of the NGC 720 study \cite{Buote02}, and therefore use halos models based on NGC 720 with a mass range of $(3-10)\times10^{12}M_{\odot}$ and ellipticities calculated within 8.5 kpc $<r<$ 14 kpc (roughly corresponding to $2.7r_e<r<4.5 r_e$ based on the effective radius given in table \ref{tab:gal_sample}). The resulting ellipticities of the simulated halos were then compared with the observed ellipticities of NGC 720 \cite{Buote02}. It was found that the NGC 720 DM halo ellipticity ($\epsilon \approx 0.35-0.4$) was consistent with both the CDM and 0.1 cm$^2$/g interaction regimes (cf. Figure 8  of \cite{Peter2013}). However the distribution in ellipticity of the simulated halos exhibited considerable scatter, making concrete conclusions based on the one observed galaxy difficult. With the supplement provided in this analysis we can consider a similar comparison, though now with a sample of galaxies rather than the singular case of NGC 720 from Ref. \cite{Buote02}. 

In figure \ref{fig:ellip_vary2} we show the normalized distribution of halo ellipticities of our sample in the background-limited and range-restricted scenarios for an NFW profile overlaid with the results of the simulations from \cite{Peter2013}. For the background limited setup it can be seen the halo ellipticities of our sample are consistent with some non-negligible DM interactions. Particularly for the Chandra data, the distribution of ellipticities is in strong agreement with  $\sigma/m =0.1$ cm$^2$/g, and the Buote et. al. (2002) \cite{Buote02} analysis of NGC 720 aligns close to the peak of the distribution. The ellipticities taken from the XMM data tend toward lower ellipticities, though also having greater spread and essentially overlapping to some degree the histogram for each interaction strength. This may to some extent be attributable to the larger PSF of XMM producing a somewhat more ``smoothed'' image compared to the Chandra data, although further investigation would be needed to determine whether this is a meaningful effect. The distributions for the range restricted case where we fit over $r_e\rightarrow 5r_e$ tend to favor more spherical halos, but also have much more scatter than the background limited radial range. 

\begin{figure*}
\begin{floatrow}
\capbtabbox{%
  \def\arraystretch{1.25}
\setlength{\tabcolsep}{5pt}
\begin{tabular}{l|cc|cc}
\hline\hline
\multicolumn{5}{c}{$\chi^2_{red.}$}\\
\hline
&\multicolumn{2}{c}{BG Limited}&\multicolumn{2}{c}{$r_e \rightarrow 5r_e$}\\
\hline
$\sigma/m_{\chi}$ & Chandra & XMM& Chandra & XMM\\
\hline
0.0 & 3.80&5.87& 3.77&7.33\\
0.1 &1.98& 4.07& 4.06&3.62\\
1.0&5.44&4.78& 4.46 & 3.25\\
\hline\hline
\end{tabular}\label{tab:twoSamp}
}{\caption{Reduced $\chi^2$ values for each of the distributions shown in the left panel of figure \ref{fig:ellip_vary2}.}}
\ffigbox{%
 \includegraphics[width=0.85\linewidth]{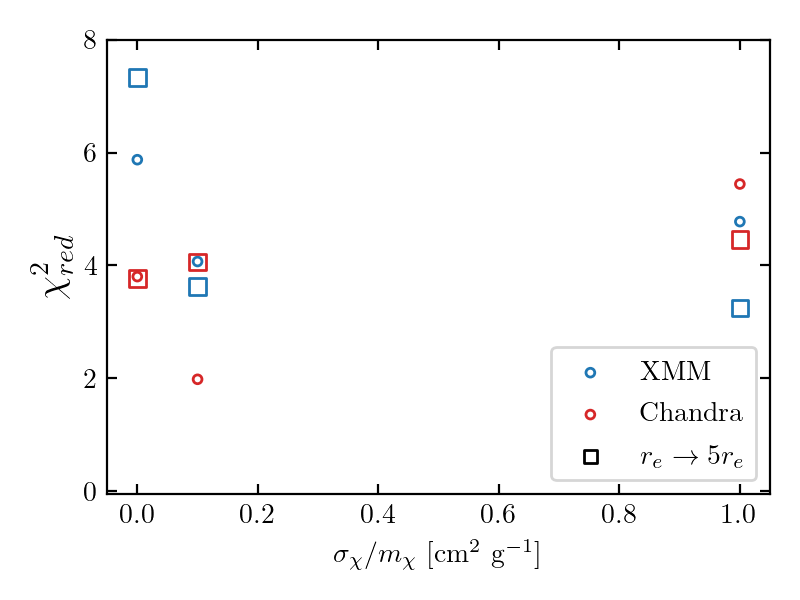}
}{%
  \caption{Visual representation of the data in table \ref{tab:twoSamp}. Colors represent results from XMM (blue) and Chandra (red) data, while the circles and squares represent the BG limited and range-restricted setups respectively.\label{fig:twoSamp}}%
}
\end{floatrow}
\end{figure*}

\begin{figure*}
\centering
    \includegraphics[width=0.46\linewidth]{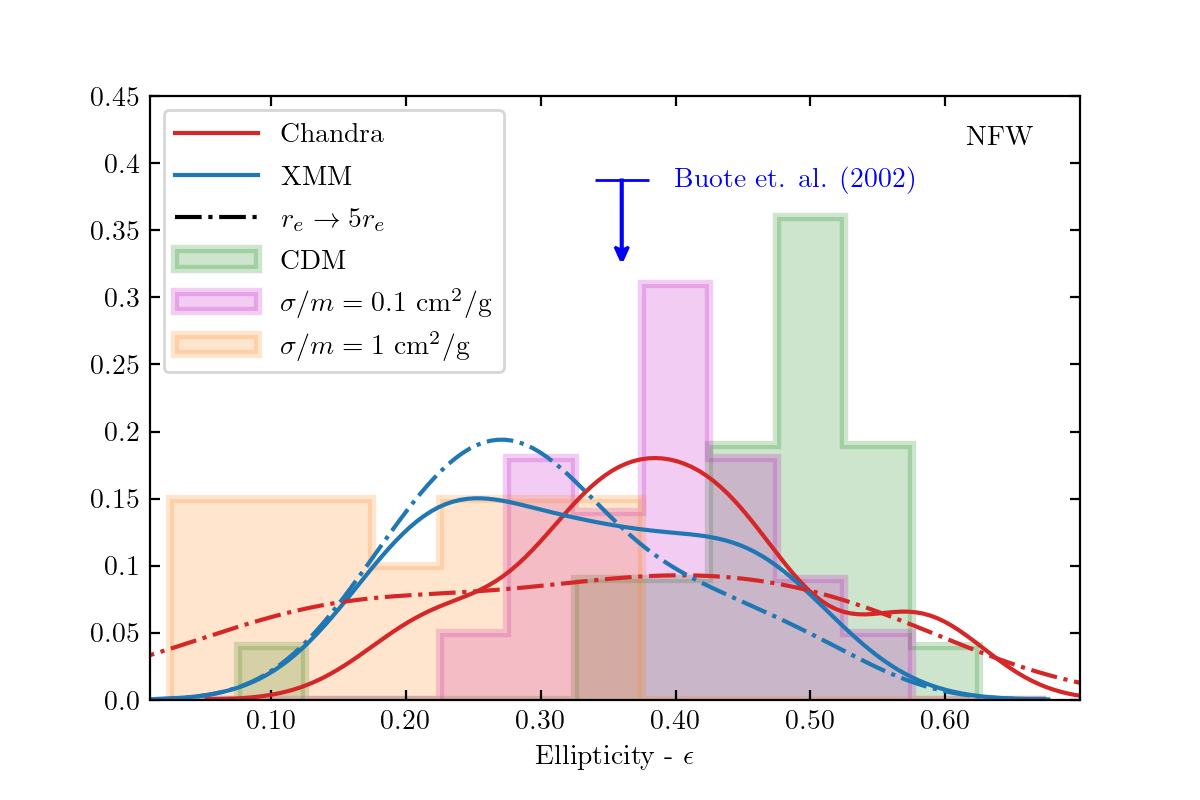}
    \includegraphics[width=0.46\linewidth]{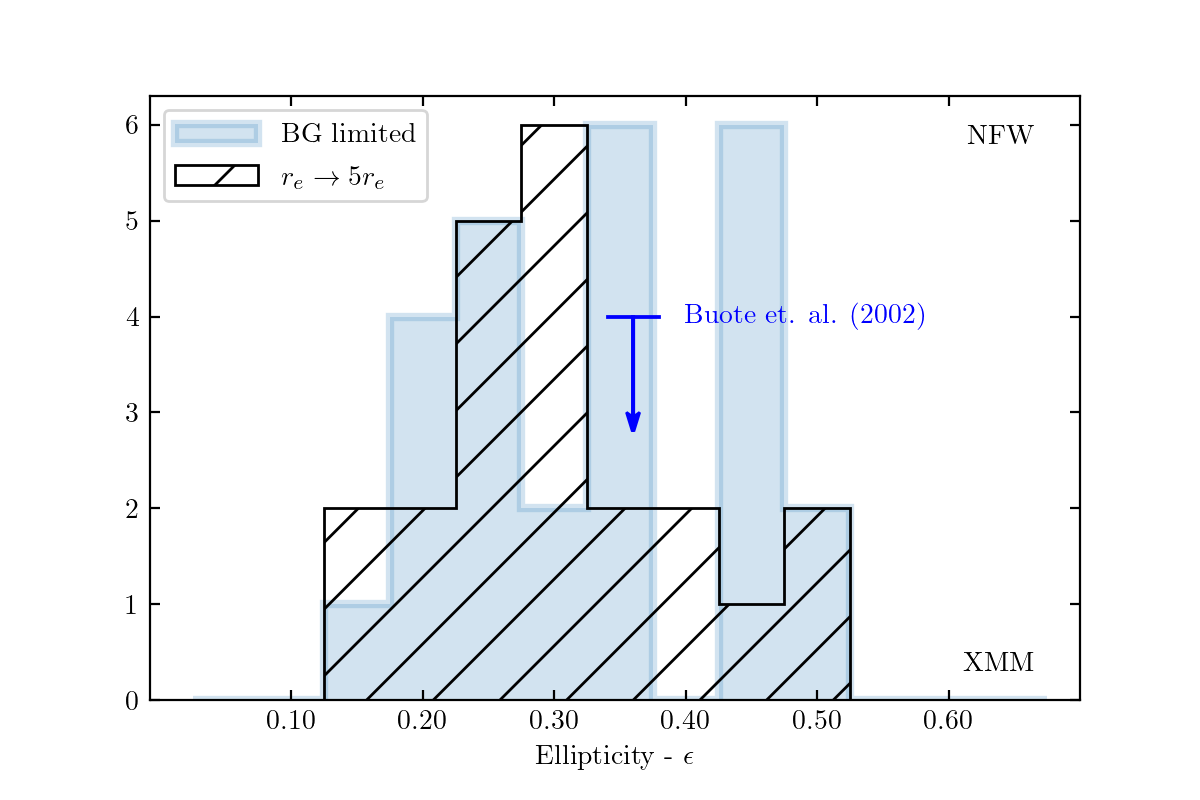}
\caption[Distribution of NFW Ellipticities Compared with Simulated Results for Varied Radial Ranges]{{\it (Left)} Normalized distributions of the best fit galaxy ellipticities from the Chandra (red) and XMM (blue) data for an NFW profile. Solid lines are for the BG limited results, while dash-dot lines are for the range-restricted results. The histograms show the simulated ellipticities for various interaction cross sections from Figure 8 of \cite{Peter2013}. We also show for reference the best-fit model and errors for the analysis of NGC 720 preformed in Ref. \cite{Buote02}. {\it (Right)} Comparison of the ellipticity histograms between the BG limited results and the range restricted $(r_e\rightarrow 5r_e)$ results. These include the results for the NFW halos from the XMM observations (i.e. the larger sample).}\label{fig:ellip_vary2} 
\end{figure*}

To quantify the level of agreement between the results of our fits and the simulated data we perform a two sample $\chi^2$ test \cite{numericalRecipes} between each of the distributions. The results are recorded in Table \ref{tab:twoSamp} and plotted in Figure \ref{fig:twoSamp}. Here we see that generally the results tend to favor some degree of self-interaction, with the lowest $\chi^2_{red}$ for the full sample corresponding to $\sigma/m =  0.1 $ cm$^2$/g. 

The sample of measured ellipticities can be applied to comparisons between additional existing and future halo simulations, and can additionally be used to constrain a variety of SIDM models. To illustrate the utility of these results in constraining specific dark matter models, we follow the theoretical framework of \cite{MCDA, feng09} and consider a dark matter candidate governed by a dark sector Coulomb interaction. We constrain the dark coupling constant $(\alpha_{\chi})$ and DM particle mass $(m_{\chi})$ by enforcing that the relaxation time be greater $\tau_{r} >10$ Gyr. The relaxation time is given in \cite{feng09} by the expression

\begin{equation}
\tau_r = E_k/\dot{E_k}= \simeq \frac{m_{\chi}^3v_0^3}{4\sqrt{\pi} \alpha_{\chi}^2\rho_{\chi}}\left(\ln \left(\frac{(b_{max} m_{\chi}v_0^2\alpha_{\chi}^{-1})^2 +1}{2}  \right)\right)^{-1},
\end{equation}
where $b_{max}$ is the impact parameter and $v_0$ corresponds to the velocity dispersion of the dark matter halo. We can rewrite this expression as:
\begin{equation}\label{eq:tau_cond}
\tau_r =  \frac{m_{\chi}^3v_0^3}{4\sqrt{\pi} \alpha_{\chi}^2\rho_{\chi}\ln\Lambda},
\end{equation}
where the ``Coulomb logarithm ‘’ has a value of $\ln \Lambda \sim 90$. In Ref. \cite{MCDA} the ellipticity of the DM halos is characterized as a function of time with the following form:
\begin{equation}\label{eq:ellip_t}
\epsilon(t) = \frac{\epsilon_0}{\frac{t}{\tau_r}\epsilon_0 + 1}
\end{equation}
Combining Eq \ref{eq:tau_cond} and Eq \ref{eq:ellip_t}, we determine a constraining relation between the dark sector coupling and DM mass given by:
\begin{equation}
\alpha_{\chi}^2 =\frac{m_{\chi}^3v_0^3}{\sqrt{\pi} \rho_{\chi} \ln \Lambda}\frac{\frac{\epsilon_0}{\epsilon} - 1}{t\epsilon_0}
\end{equation}
We adopt nominal values for the velocity dispersion and density profiles of 250 km/s and 1 GeV/cm$^3$ over the spatial extent of the region in which we perform our fits --- essentially the region where the dark matter dominates the mass contribution but that still provides reliable x-ray data. The choice of these values is based on the results of mass and concentration parameter measurements in NGC 720 (see e.g. \cite{720_mass,chandra_galaxy_selection}) and are taken to be roughly representative of our sample. In future works, modeling each of the galaxies in a more detailed manner could provide slightly different results.

\begin{figure*}
\centering
\includegraphics[width=0.75\linewidth]{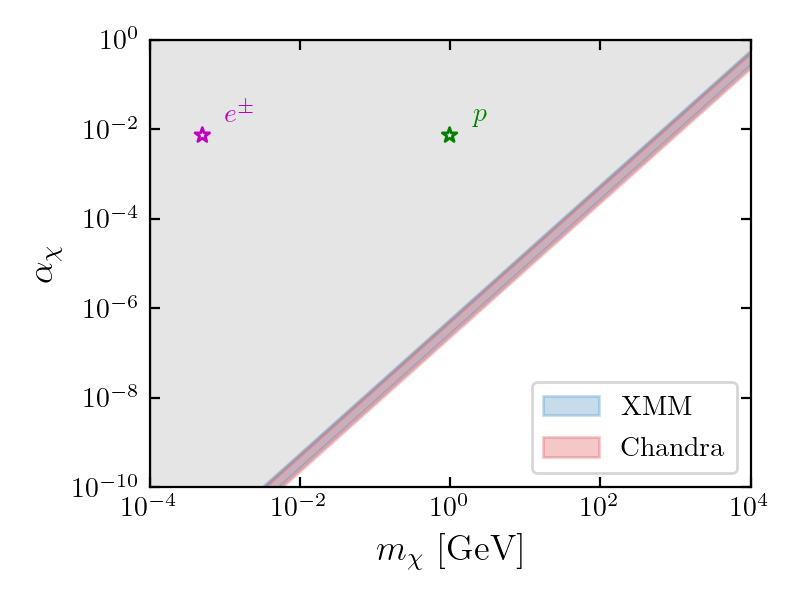}
\caption[Constraints on the $\alpha_{\chi}-m_{\chi}$ parameter space]{Constraints on the $\alpha_{\chi}-m_{\chi}$  space from our sample of galaxies ellipticities. The excluded region is shown in grey, and the blue and red bands give the range of limits from the combined results of the XMM and Chandra data respectively. Although the two bands mostly overlap, the upper and lower limits of the XMM band (blue) are slightly higher than those of the Chandra band (red). We also show for reference where the electron (magenta star) and proton (green star) would be located on a Standard Model version of this plot.}
\label{fig:alphaVm}
\end{figure*}

The resulting combined constraints using the measured ellipticities are shown in figure \ref{fig:alphaVm} for both the Chandra and XMM data sets. The constraints are quite similar between the two and mostly overlap. The XMM data prefers slightly weaker limits, likely attributable to the preference for rounder halos found in the fits (see figure \ref{fig:ellip_vary2} and the related discussion). Ultimately our constraints found here are in good agreement with those of Ref. \cite{feng09}, with the benefit in our work of having used a full sample of measured ellipticities rather than the sole measurements of NGC 720.

\section{Conclusions}\label{sec:conclusion}
In this analysis we have studied a sample of isolated elliptical galaxies as a probe of the dark matter halo shape. Using archival data from the XMM-Newton and Chandra telescopes we analyzed the X-ray surface brightness and ellipticity profiles of 11 elliptical galaxies. By selecting galaxies that meet the criteria of being relaxed, isolated, and approximately isothermal, we made the assumption plausibly well-justified assumption of hydrostatic equilibrium with the gravitational potential. Under this assumption we showed the relation between the shape of the 3D matter distribution and the 2D projected X-ray image, which allowed us to model the 3D matter distribution and fit to the data, thereby determining the underlying shape for the assumed matter distribution.

We considered three separate spheroidal mass distributions including an NFW, Hernquist, and pseudo-isothermal profile. Additionally, we considered both prolate and oblate configurations for each profile, effectively bracketing the range of triaxial models. The best fitting ellipticities did not appear to have a significant relation to the profile. We also considered two choices of radial range over which we performed our fits: one that extended until a chip edge was reached or the background became dominant, and another range restricted setup motivated by previous studies and simulations. While the range restricted setup generally favored lower ellipticities and had higher scatter, the differences between the two choices do not significantly alter our conclusions.

For most of the galaxies in the sample the ellipticities fell roughly within the range of $\epsilon\approx 0.2-0.5$. Comparing the measured ellipticities with simulations of DM halos that have varying degrees of self-interactions shows consistency and a marginal statistical perference with interaction cross-sections of $\sigma/m = 0.1$ cm$^2$/g. However, the simulations of Ref. \cite{Peter2013} used for this comparison reported significant scatter. This is also apparent in our observational results, and the ellipticity distribution also overlaps significantly with the distribution for CDM halos. We note that these findings are highly consistent with the comparison between the simulations and the ellipticity analysis of NGC 720 performed in Ref. \cite{Buote02}. 

While the work of Ref. \cite{Buote02} and subsequent studies have shown that X-ray shapes of elliptical galaxies can have powerful constraining ability on SIDM, most of these analyses have used only the singular case of NGC 720. The results presented here now allow for a comparison using a statistically meaningful sample in order to produce constraints as demonstrated in section \ref{sec:sidm}. While these results can be used constrain specific models, our sample also now allows  detailed comparison with simulations containing only DM as well as future simulations that also seek to incorporate baryonic effects.

\acknowledgments
This work is funded by NASA award number 80NSSC20K0434. A.M. is partially supported by a Department of Education GAANN fellowship. 
%\end{acknowledgments}
%\bibliographystyle{plainnat}
%\bibliographystyle{natbib}
\bibliographystyle{unsrtnat}
\bibliography{ref}

\begin{thebibliography}{81}
\providecommand{\natexlab}[1]{#1}
\providecommand{\url}[1]{\texttt{#1}}
\expandafter\ifx\csname urlstyle\endcsname\relax
  \providecommand{\doi}[1]{doi: #1}\else
  \providecommand{\doi}{doi: \begingroup \urlstyle{rm}\Url}\fi

\bibitem[{Springel} et~al.(2006){Springel}, {Frenk}, and {White}]{LSS}
Volker {Springel}, Carlos~S. {Frenk}, and Simon D.~M. {White}.
\newblock {The large-scale structure of the Universe}.
\newblock \emph{\nat}, 440\penalty0 (7088):\penalty0 1137--1144, April 2006.
\newblock \doi{10.1038/nature04805}.

\bibitem[Arcadi et~al.(2018)Arcadi, Dutra, Ghosh, Lindner, Mambrini, Pierre,
  Profumo, and Queiroz]{Arcadi:2017kky}
Giorgio Arcadi, Ma\'\i{}ra Dutra, Pradipta Ghosh, Manfred Lindner, Yann
  Mambrini, Mathias Pierre, Stefano Profumo, and Farinaldo~S. Queiroz.
\newblock {The waning of the WIMP? A review of models, searches, and
  constraints}.
\newblock \emph{Eur. Phys. J. C}, 78\penalty0 (3):\penalty0 203, 2018.
\newblock \doi{10.1140/epjc/s10052-018-5662-y}.

\bibitem[{Klypin} et~al.(1999){Klypin}, {Kravtsov}, {Valenzuela}, and
  {Prada}]{KlypinMS}
Anatoly {Klypin}, Andrey~V. {Kravtsov}, Octavio {Valenzuela}, and Francisco
  {Prada}.
\newblock {Where Are the Missing Galactic Satellites?}
\newblock \emph{\apj}, 522\penalty0 (1):\penalty0 82--92, September 1999.
\newblock \doi{10.1086/307643}.

\bibitem[{Moore} et~al.(1999){Moore}, {Ghigna}, {Governato}, {Lake}, {Quinn},
  {Stadel}, and {Tozzi}]{mooreMS}
Ben {Moore}, Sebastiano {Ghigna}, Fabio {Governato}, George {Lake}, Thomas
  {Quinn}, Joachim {Stadel}, and Paolo {Tozzi}.
\newblock {Dark Matter Substructure within Galactic Halos}.
\newblock \emph{\apjl}, 524\penalty0 (1):\penalty0 L19--L22, October 1999.
\newblock \doi{10.1086/312287}.

\bibitem[{Boylan-Kolchin} et~al.(2011){Boylan-Kolchin}, {Bullock}, and
  {Kaplinghat}]{TBTF2011}
Michael {Boylan-Kolchin}, James~S. {Bullock}, and Manoj {Kaplinghat}.
\newblock {Too big to fail? The puzzling darkness of massive Milky Way
  subhaloes}.
\newblock \emph{\mnras}, 415\penalty0 (1):\penalty0 L40--L44, July 2011.
\newblock \doi{10.1111/j.1745-3933.2011.01074.x}.

\bibitem[{Boylan-Kolchin} et~al.(2012){Boylan-Kolchin}, {Bullock}, and
  {Kaplinghat}]{TBTF2012}
Michael {Boylan-Kolchin}, James~S. {Bullock}, and Manoj {Kaplinghat}.
\newblock {The Milky Way's bright satellites as an apparent failure of
  {\ensuremath{\Lambda}}CDM}.
\newblock \emph{\mnras}, 422\penalty0 (2):\penalty0 1203--1218, May 2012.
\newblock \doi{10.1111/j.1365-2966.2012.20695.x}.

\bibitem[{Flores} and {Primack}(1994)]{FloresPrimack}
Ricardo~A. {Flores} and Joel~R. {Primack}.
\newblock {Observational and Theoretical Constraints on Singular Dark Matter
  Halos}.
\newblock \emph{\apjl}, 427:\penalty0 L1, May 1994.
\newblock \doi{10.1086/187350}.

\bibitem[{Moore}(1994)]{mooreCC}
Ben {Moore}.
\newblock {Evidence against dissipation-less dark matter from observations of
  galaxy haloes}.
\newblock \emph{\nat}, 370\penalty0 (6491):\penalty0 629--631, August 1994.
\newblock \doi{10.1038/370629a0}.

\bibitem[{Kamada} et~al.(2017){Kamada}, {Kaplinghat}, {Pace}, and
  {Yu}]{KamadaCC}
Ayuki {Kamada}, Manoj {Kaplinghat}, Andrew~B. {Pace}, and Hai-Bo {Yu}.
\newblock {Self-Interacting Dark Matter Can Explain Diverse Galactic Rotation
  Curves}.
\newblock \emph{\prl}, 119\penalty0 (11):\penalty0 111102, September 2017.
\newblock \doi{10.1103/PhysRevLett.119.111102}.

\bibitem[{Primack}(2001)]{primackNature}
Joel~R. {Primack}.
\newblock {The Nature of Dark Matter}.
\newblock \emph{arXiv e-prints}, art. astro-ph/0112255, December 2001.

\bibitem[{Navarro} et~al.(1996){Navarro}, {Eke}, and {Frenk}]{navarroDwarfs}
Julio~F. {Navarro}, Vincent~R. {Eke}, and Carlos~S. {Frenk}.
\newblock {The cores of dwarf galaxy haloes}.
\newblock \emph{\mnras}, 283\penalty0 (3):\penalty0 L72--L78, December 1996.
\newblock \doi{10.1093/mnras/283.3.L72}.

\bibitem[{Governato} et~al.(2012){Governato}, {Zolotov}, {Pontzen},
  {Christensen}, {Oh}, {Brooks}, {Quinn}, {Shen}, and
  {Wadsley}]{governatoCusps}
F.~{Governato}, A.~{Zolotov}, A.~{Pontzen}, C.~{Christensen}, S.~H. {Oh}, A.~M.
  {Brooks}, T.~{Quinn}, S.~{Shen}, and J.~{Wadsley}.
\newblock {Cuspy no more: how outflows affect the central dark matter and
  baryon distribution in {\ensuremath{\Lambda}} cold dark matter galaxies}.
\newblock \emph{\mnras}, 422\penalty0 (2):\penalty0 1231--1240, May 2012.
\newblock \doi{10.1111/j.1365-2966.2012.20696.x}.

\bibitem[{Brooks} and {Zolotov}(2014)]{brooksDwarfs}
Alyson~M. {Brooks} and Adi {Zolotov}.
\newblock {Why Baryons Matter: The Kinematics of Dwarf Spheroidal Satellites}.
\newblock \emph{\apj}, 786\penalty0 (2):\penalty0 87, May 2014.
\newblock \doi{10.1088/0004-637X/786/2/87}.

\bibitem[{Zolotov} et~al.(2012){Zolotov}, {Brooks}, {Willman}, {Governato},
  {Pontzen}, {Christensen}, {Dekel}, {Quinn}, {Shen}, and
  {Wadsley}]{ZolotovBrooks}
Adi {Zolotov}, Alyson~M. {Brooks}, Beth {Willman}, Fabio {Governato}, Andrew
  {Pontzen}, Charlotte {Christensen}, Avishai {Dekel}, Tom {Quinn}, Sijing
  {Shen}, and James {Wadsley}.
\newblock {Baryons Matter: Why Luminous Satellite Galaxies have Reduced Central
  Masses}.
\newblock \emph{\apj}, 761\penalty0 (1):\penalty0 71, December 2012.
\newblock \doi{10.1088/0004-637X/761/1/71}.

\bibitem[{Bullock} and {Boylan-Kolchin}(2017)]{bullockReview}
James~S. {Bullock} and Michael {Boylan-Kolchin}.
\newblock {Small-Scale Challenges to the {\ensuremath{\Lambda}}CDM Paradigm}.
\newblock \emph{\araa}, 55\penalty0 (1):\penalty0 343--387, August 2017.
\newblock \doi{10.1146/annurev-astro-091916-055313}.

\bibitem[{Governato} et~al.(2010){Governato}, {Brook}, {Mayer}, {Brooks},
  {Rhee}, {Wadsley}, {Jonsson}, {Willman}, {Stinson}, {Quinn}, and
  {Madau}]{governatoDwarfs}
F.~{Governato}, C.~{Brook}, L.~{Mayer}, A.~{Brooks}, G.~{Rhee}, J.~{Wadsley},
  P.~{Jonsson}, B.~{Willman}, G.~{Stinson}, T.~{Quinn}, and P.~{Madau}.
\newblock {Bulgeless dwarf galaxies and dark matter cores from supernova-driven
  outflows}.
\newblock \emph{\nat}, 463\penalty0 (7278):\penalty0 203--206, January 2010.
\newblock \doi{10.1038/nature08640}.

\bibitem[{Amorisco} et~al.(2014){Amorisco}, {Zavala}, and {de Boer}]{amorisco}
N.~C. {Amorisco}, J.~{Zavala}, and T.~J.~L. {de Boer}.
\newblock {Dark Matter Cores in the Fornax and Sculptor Dwarf Galaxies: Joining
  Halo Assembly and Detailed Star Formation Histories}.
\newblock \emph{\apjl}, 782\penalty0 (2):\penalty0 L39, February 2014.
\newblock \doi{10.1088/2041-8205/782/2/L39}.

\bibitem[{Kirby} et~al.(2014){Kirby}, {Bullock}, {Boylan-Kolchin},
  {Kaplinghat}, and {Cohen}]{kirby2014}
Evan~N. {Kirby}, James~S. {Bullock}, Michael {Boylan-Kolchin}, Manoj
  {Kaplinghat}, and Judith~G. {Cohen}.
\newblock {The dynamics of isolated Local Group galaxies}.
\newblock \emph{\mnras}, 439\penalty0 (1):\penalty0 1015--1027, March 2014.
\newblock \doi{10.1093/mnras/stu025}.

\bibitem[{Kaplinghat} et~al.(2014){Kaplinghat}, {Keeley}, {Linden}, and
  {Yu}]{kapKeeleyLinden2014}
Manoj {Kaplinghat}, Ryan~E. {Keeley}, Tim {Linden}, and Hai-Bo {Yu}.
\newblock {Tying Dark Matter to Baryons with Self-Interactions}.
\newblock \emph{\prl}, 113\penalty0 (2):\penalty0 021302, July 2014.
\newblock \doi{10.1103/PhysRevLett.113.021302}.

\bibitem[{Papastergis} and {Ponomareva}(2017)]{PapPon2017}
E.~{Papastergis} and A.~A. {Ponomareva}.
\newblock {Testing core creation in hydrodynamical simulations using the HI
  kinematics of field dwarfs}.
\newblock \emph{\aap}, 601:\penalty0 A1, May 2017.
\newblock \doi{10.1051/0004-6361/201629546}.

\bibitem[{Spergel} and {Steinhardt}(2000)]{SS}
David~N. {Spergel} and Paul~J. {Steinhardt}.
\newblock {Observational Evidence for Self-Interacting Cold Dark Matter}.
\newblock \emph{\prl}, 84\penalty0 (17):\penalty0 3760--3763, April 2000.
\newblock \doi{10.1103/PhysRevLett.84.3760}.

\bibitem[{Tulin} and {Yu}(2018)]{tulinYuReview}
Sean {Tulin} and Hai-Bo {Yu}.
\newblock {Dark matter self-interactions and small scale structure}.
\newblock \emph{\physrep}, 730:\penalty0 1--57, February 2018.
\newblock \doi{10.1016/j.physrep.2017.11.004}.

\bibitem[{Vogelsberger} et~al.(2012){Vogelsberger}, {Zavala}, and
  {Loeb}]{VZL2012}
Mark {Vogelsberger}, Jesus {Zavala}, and Abraham {Loeb}.
\newblock {Subhaloes in self-interacting galactic dark matter haloes}.
\newblock \emph{\mnras}, 423\penalty0 (4):\penalty0 3740--3752, July 2012.
\newblock \doi{10.1111/j.1365-2966.2012.21182.x}.

\bibitem[{Peter} et~al.(2013){Peter}, {Rocha}, {Bullock}, and
  {Kaplinghat}]{Peter2013}
Annika H.~G. {Peter}, Miguel {Rocha}, James~S. {Bullock}, and Manoj
  {Kaplinghat}.
\newblock {Cosmological simulations with self-interacting dark matter - II.
  Halo shapes versus observations}.
\newblock \emph{\mnras}, 430\penalty0 (1):\penalty0 105--120, March 2013.
\newblock \doi{10.1093/mnras/sts535}.

\bibitem[{Elbert} et~al.(2015){Elbert}, {Bullock}, {Garrison-Kimmel}, {Rocha},
  {O{\~n}orbe}, and {Peter}]{Elbert2015}
Oliver~D. {Elbert}, James~S. {Bullock}, Shea {Garrison-Kimmel}, Miguel {Rocha},
  Jose {O{\~n}orbe}, and Annika H.~G. {Peter}.
\newblock {Core formation in dwarf haloes with self-interacting dark matter: no
  fine-tuning necessary}.
\newblock \emph{\mnras}, 453\penalty0 (1):\penalty0 29--37, October 2015.
\newblock \doi{10.1093/mnras/stv1470}.

\bibitem[{Dav{\'e}} et~al.(2001){Dav{\'e}}, {Spergel}, {Steinhardt}, and
  {Wandelt}]{DSS2000}
Romeel {Dav{\'e}}, David~N. {Spergel}, Paul~J. {Steinhardt}, and Benjamin~D.
  {Wandelt}.
\newblock {Halo Properties in Cosmological Simulations of Self-interacting Cold
  Dark Matter}.
\newblock \emph{\apj}, 547\penalty0 (2):\penalty0 574--589, February 2001.
\newblock \doi{10.1086/318417}.

\bibitem[{Miralda-Escud{\'e}}(2002)]{ME02}
Jordi {Miralda-Escud{\'e}}.
\newblock {A Test of the Collisional Dark Matter Hypothesis from Cluster
  Lensing}.
\newblock \emph{\apj}, 564\penalty0 (1):\penalty0 60--64, January 2002.
\newblock \doi{10.1086/324138}.

\bibitem[{Andrade} et~al.(2019){Andrade}, {Minor}, {Nierenberg}, and
  {Kaplinghat}]{andrade}
Kevin~E. {Andrade}, Quinn {Minor}, Anna {Nierenberg}, and Manoj {Kaplinghat}.
\newblock {Detecting dark matter cores in galaxy clusters with strong lensing}.
\newblock \emph{\mnras}, 487\penalty0 (2):\penalty0 1905--1926, August 2019.
\newblock \doi{10.1093/mnras/stz1360}.

\bibitem[{Randall} et~al.(2008){Randall}, {Markevitch}, {Clowe}, {Gonzalez},
  and {Brada{\v{c}}}]{randallBullet}
Scott~W. {Randall}, Maxim {Markevitch}, Douglas {Clowe}, Anthony~H. {Gonzalez},
  and Marusa {Brada{\v{c}}}.
\newblock {Constraints on the Self-Interaction Cross Section of Dark Matter
  from Numerical Simulations of the Merging Galaxy Cluster 1E 0657-56}.
\newblock \emph{\apj}, 679\penalty0 (2):\penalty0 1173--1180, June 2008.
\newblock \doi{10.1086/587859}.

\bibitem[{Markevitch} et~al.(2004){Markevitch}, {Gonzalez}, {Clowe},
  {Vikhlinin}, {Forman}, {Jones}, {Murray}, and {Tucker}]{M04Bullet}
M.~{Markevitch}, A.~H. {Gonzalez}, D.~{Clowe}, A.~{Vikhlinin}, W.~{Forman},
  C.~{Jones}, S.~{Murray}, and W.~{Tucker}.
\newblock {Direct Constraints on the Dark Matter Self-Interaction Cross Section
  from the Merging Galaxy Cluster 1E 0657-56}.
\newblock \emph{\apj}, 606\penalty0 (2):\penalty0 819--824, May 2004.
\newblock \doi{10.1086/383178}.

\bibitem[{Buote} et~al.(2002){Buote}, {Jeltema}, {Canizares}, and
  {Garmire}]{Buote02}
David~A. {Buote}, Tesla~E. {Jeltema}, Claude~R. {Canizares}, and Gordon~P.
  {Garmire}.
\newblock {Chandra Evidence of a Flattened, Triaxial Dark Matter Halo in the
  Elliptical Galaxy NGC 720}.
\newblock \emph{\apj}, 577\penalty0 (1):\penalty0 183--196, September 2002.
\newblock \doi{10.1086/342158}.

\bibitem[{Rocha} et~al.(2013){Rocha}, {Peter}, {Bullock}, {Kaplinghat},
  {Garrison-Kimmel}, {O{\~n}orbe}, and {Moustakas}]{Rocha}
Miguel {Rocha}, Annika H.~G. {Peter}, James~S. {Bullock}, Manoj {Kaplinghat},
  Shea {Garrison-Kimmel}, Jose {O{\~n}orbe}, and Leonidas~A. {Moustakas}.
\newblock {Cosmological simulations with self-interacting dark matter - I.
  Constant-density cores and substructure}.
\newblock \emph{\mnras}, 430\penalty0 (1):\penalty0 81--104, March 2013.
\newblock \doi{10.1093/mnras/sts514}.

\bibitem[{Kaplinghat} et~al.(2016){Kaplinghat}, {Tulin}, and
  {Yu}]{SIDMVelocity}
Manoj {Kaplinghat}, Sean {Tulin}, and Hai-Bo {Yu}.
\newblock {Dark Matter Halos as Particle Colliders: Unified Solution to
  Small-Scale Structure Puzzles from Dwarfs to Clusters}.
\newblock \emph{\prl}, 116\penalty0 (4):\penalty0 041302, January 2016.
\newblock \doi{10.1103/PhysRevLett.116.041302}.

\bibitem[{Feng} et~al.(2010){Feng}, {Kaplinghat}, and {Yu}]{fengKap}
Jonathan~L. {Feng}, Manoj {Kaplinghat}, and Hai-Bo {Yu}.
\newblock {Halo-Shape and Relic-Density Exclusions of Sommerfeld-Enhanced Dark
  Matter Explanations of Cosmic Ray Excesses}.
\newblock \emph{\prl}, 104\penalty0 (15):\penalty0 151301, April 2010.
\newblock \doi{10.1103/PhysRevLett.104.151301}.

\bibitem[{Loeb} and {Weiner}(2011)]{loeb}
Abraham {Loeb} and Neal {Weiner}.
\newblock {Cores in Dwarf Galaxies from Dark Matter with a Yukawa Potential}.
\newblock \emph{\prl}, 106\penalty0 (17):\penalty0 171302, April 2011.
\newblock \doi{10.1103/PhysRevLett.106.171302}.

\bibitem[{Buckley} and {Fox}(2010)]{buckley}
Matthew~R. {Buckley} and Patrick~J. {Fox}.
\newblock {Dark matter self-interactions and light force carriers}.
\newblock \emph{\prd}, 81\penalty0 (8):\penalty0 083522, April 2010.
\newblock \doi{10.1103/PhysRevD.81.083522}.

\bibitem[{Tulin} et~al.(2013{\natexlab{a}}){Tulin}, {Yu}, and {Zurek}]{TYZ1}
Sean {Tulin}, Hai-Bo {Yu}, and Kathryn~M. {Zurek}.
\newblock {Resonant Dark Forces and Small-Scale Structure}.
\newblock \emph{\prl}, 110\penalty0 (11):\penalty0 111301, March
  2013{\natexlab{a}}.
\newblock \doi{10.1103/PhysRevLett.110.111301}.

\bibitem[{Tulin} et~al.(2013{\natexlab{b}}){Tulin}, {Yu}, and {Zurek}]{TYZ2}
Sean {Tulin}, Hai-Bo {Yu}, and Kathryn~M. {Zurek}.
\newblock {Beyond collisionless dark matter: Particle physics dynamics for dark
  matter halo structure}.
\newblock \emph{\prd}, 87\penalty0 (11):\penalty0 115007, June
  2013{\natexlab{b}}.
\newblock \doi{10.1103/PhysRevD.87.115007}.

\bibitem[{Buote} and {Humphrey}(2012{\natexlab{a}})]{BH_review}
David~A. {Buote} and Philip~J. {Humphrey}.
\newblock \emph{{Dark Matter in Elliptical Galaxies}}, volume 378, page 235.
\newblock 2012{\natexlab{a}}.
\newblock \doi{10.1007/978-1-4614-0580-1_8}.

\bibitem[{Binney} and {Strimpel}(1978)]{BinneyStrimpel}
James {Binney} and Oliver {Strimpel}.
\newblock {Predicting the X-ray brightness distributions of cluster sources -
  1. Estimating the potentials}.
\newblock \emph{\mnras}, 185:\penalty0 473--484, November 1978.
\newblock \doi{10.1093/mnras/185.3.473}.

\bibitem[{Buote} and {Canizares}(1992)]{BCAbell92}
David~A. {Buote} and Claude~R. {Canizares}.
\newblock {X-Ray Constraints on the Shape of the Dark Matter in Five Abell
  Clusters}.
\newblock \emph{\apj}, 400:\penalty0 385, December 1992.
\newblock \doi{10.1086/172004}.

\bibitem[{Buote} and {Canizares}(1994)]{buoteCanizares}
David~A. {Buote} and Claude~R. {Canizares}.
\newblock {Geometrical Evidence for Dark Matter: X-Ray Constraints on the Mass
  of the Elliptical Galaxy NGC 720}.
\newblock \emph{\apj}, 427:\penalty0 86, May 1994.
\newblock \doi{10.1086/174123}.

\bibitem[{Buote} and {Canizares}(1996{\natexlab{a}})]{BC_NGC1332_96}
David~A. {Buote} and Claude~R. {Canizares}.
\newblock {X-Ray Constraints on the Intrinsic Shape of the Lenticular Galaxy
  NGC 1332}.
\newblock \emph{\apj}, 457:\penalty0 177, January 1996{\natexlab{a}}.
\newblock \doi{10.1086/176721}.

\bibitem[{Buote} and {Canizares}(1998)]{BC_NGC3923_98}
David~A. {Buote} and Claude~R. {Canizares}.
\newblock {X-ray isophote shapes and the mass of NGC 3923}.
\newblock \emph{\mnras}, 298\penalty0 (3):\penalty0 811--823, August 1998.
\newblock \doi{10.1046/j.1365-8711.1998.01663.x}.

\bibitem[{Cyr-Racine} and {Sigurdson}(2013)]{atomicDM}
Francis-Yan {Cyr-Racine} and Kris {Sigurdson}.
\newblock {Cosmology of atomic dark matter}.
\newblock \emph{\prd}, 87\penalty0 (10):\penalty0 103515, May 2013.
\newblock \doi{10.1103/PhysRevD.87.103515}.

\bibitem[{Fan} et~al.(2013){Fan}, {Katz}, {Randall}, and {Reece}]{fan2013}
JiJi {Fan}, Andrey {Katz}, Lisa {Randall}, and Matthew {Reece}.
\newblock {Double-Disk Dark Matter}.
\newblock \emph{Physics of the Dark Universe}, 2\penalty0 (3):\penalty0
  139--156, September 2013.
\newblock \doi{10.1016/j.dark.2013.07.001}.

\bibitem[{McDermott} et~al.(2011){McDermott}, {Yu}, and {Zurek}]{mcdermott}
Samuel~D. {McDermott}, Hai-Bo {Yu}, and Kathryn~M. {Zurek}.
\newblock {Turning off the lights: How dark is dark matter?}
\newblock \emph{\prd}, 83\penalty0 (6):\penalty0 063509, March 2011.
\newblock \doi{10.1103/PhysRevD.83.063509}.

\bibitem[{Feng} et~al.(2009){Feng}, {Kaplinghat}, {Tu}, and {Yu}]{feng09}
Jonathan~L. {Feng}, Manoj {Kaplinghat}, Huitzu {Tu}, and Hai-Bo {Yu}.
\newblock {Hidden charged dark matter}.
\newblock \emph{\jcap}, 2009\penalty0 (7):\penalty0 004, July 2009.
\newblock \doi{10.1088/1475-7516/2009/07/004}.

\bibitem[{Agrawal} et~al.(2017){Agrawal}, {Cyr-Racine}, {Randall}, and
  {Scholtz}]{MCDA}
Prateek {Agrawal}, Francis-Yan {Cyr-Racine}, Lisa {Randall}, and Jakub
  {Scholtz}.
\newblock {Make dark matter charged again}.
\newblock \emph{\jcap}, 2017\penalty0 (5):\penalty0 022, May 2017.
\newblock \doi{10.1088/1475-7516/2017/05/022}.

\bibitem[{Milgrom}(2012)]{MONDellipticals}
Mordehai {Milgrom}.
\newblock {Testing MOND over a Wide Acceleration Range in X-Ray Ellipticals}.
\newblock \emph{\prl}, 109\penalty0 (13):\penalty0 131101, September 2012.
\newblock \doi{10.1103/PhysRevLett.109.131101}.

\bibitem[Binney and Tremaine(2011)]{BT}
J.~Binney and S.~Tremaine.
\newblock \emph{Galactic Dynamics: Second Edition}.
\newblock Princeton Series in Astrophysics. Princeton University Press, 2011.
\newblock ISBN 9781400828722.
\newblock URL \url{https://books.google.com/books?id=6mF4CKxlbLsC}.

\bibitem[{Buote} and {Humphrey}(2012{\natexlab{b}})]{BuoteHumphrey2012}
David~A. {Buote} and Philip~J. {Humphrey}.
\newblock {Spherically averaging ellipsoidal galaxy clusters in X-ray and
  Sunyaev-Zel'dovich studies - II. Biases}.
\newblock \emph{\mnras}, 421\penalty0 (2):\penalty0 1399--1420, April
  2012{\natexlab{b}}.
\newblock \doi{10.1111/j.1365-2966.2011.20399.x}.

\bibitem[{Navarro} et~al.(1997){Navarro}, {Frenk}, and {White}]{NFW97}
J.~F. {Navarro}, C.~S. {Frenk}, and S.~D.~M. {White}.
\newblock {A Universal Density Profile from Hierarchical Clustering}.
\newblock \emph{\apj}, 490:\penalty0 493--508, December 1997.
\newblock \doi{10.1086/304888}.

\bibitem[{Hernquist}(1990)]{hernquist}
Lars {Hernquist}.
\newblock {An Analytical Model for Spherical Galaxies and Bulges}.
\newblock \emph{\apj}, 356:\penalty0 359, June 1990.
\newblock \doi{10.1086/168845}.

\bibitem[{Humphrey} et~al.(2006){Humphrey}, {Buote}, {Gastaldello},
  {Zappacosta}, {Bullock}, {Brighenti}, and
  {Mathews}]{chandra_galaxy_selection}
Philip~J. {Humphrey}, David~A. {Buote}, Fabio {Gastaldello}, Luca {Zappacosta},
  James~S. {Bullock}, Fabrizio {Brighenti}, and William~G. {Mathews}.
\newblock {A Chandra View of Dark Matter in Early-Type Galaxies}.
\newblock \emph{\apj}, 646\penalty0 (2):\penalty0 899--918, August 2006.
\newblock \doi{10.1086/505019}.

\bibitem[{Cavagnolo} et~al.(2010){Cavagnolo}, {McNamara}, {Nulsen}, {Carilli},
  {Jones}, and {B{\^\i}rzan}]{M84}
K.~W. {Cavagnolo}, B.~R. {McNamara}, P.~E.~J. {Nulsen}, C.~L. {Carilli},
  C.~{Jones}, and L.~{B{\^\i}rzan}.
\newblock {A Relationship Between AGN Jet Power and Radio Power}.
\newblock \emph{\apj}, 720\penalty0 (2):\penalty0 1066--1072, September 2010.
\newblock \doi{10.1088/0004-637X/720/2/1066}.

\bibitem[{Huchra} et~al.(2012){Huchra}, {Macri}, {Masters}, {Jarrett},
  {Berlind}, {Calkins}, {Crook}, {Cutri}, {Erdo{\v{g}}du}, {Falco}, {George},
  {Hutcheson}, {Lahav}, {Mader}, {Mink}, {Martimbeau}, {Schneider},
  {Skrutskie}, {Tokarz}, and {Westover}]{2mass}
John~P. {Huchra}, Lucas~M. {Macri}, Karen~L. {Masters}, Thomas~H. {Jarrett},
  Perry {Berlind}, Michael {Calkins}, Aidan~C. {Crook}, Roc {Cutri}, Pirin
  {Erdo{\v{g}}du}, Emilio {Falco}, Teddy {George}, Conrad~M. {Hutcheson}, Ofer
  {Lahav}, Jeff {Mader}, Jessica~D. {Mink}, Nathalie {Martimbeau}, Stephen
  {Schneider}, Michael {Skrutskie}, Susan {Tokarz}, and Michael {Westover}.
\newblock {The 2MASS Redshift Survey{\textemdash}Description and Data Release}.
\newblock \emph{\apjs}, 199\penalty0 (2):\penalty0 26, April 2012.
\newblock \doi{10.1088/0067-0049/199/2/26}.

\bibitem[{Springob} et~al.(2014){Springob}, {Magoulas}, {Colless}, {Mould},
  {Erdo{\u{g}}du}, {Jones}, {Lucey}, {Campbell}, and {Fluke}]{springbob}
Christopher~M. {Springob}, Christina {Magoulas}, Matthew {Colless}, Jeremy
  {Mould}, Pirin {Erdo{\u{g}}du}, D.~Heath {Jones}, John~R. {Lucey}, Lachlan
  {Campbell}, and Christopher~J. {Fluke}.
\newblock {The 6dF Galaxy Survey: peculiar velocity field and cosmography}.
\newblock \emph{\mnras}, 445\penalty0 (3):\penalty0 2677--2697, December 2014.
\newblock \doi{10.1093/mnras/stu1743}.

\bibitem[{Theureau} et~al.(2007){Theureau}, {Hanski}, {Coudreau}, {Hallet}, and
  {Martin}]{tf_dist}
G.~{Theureau}, M.~O. {Hanski}, N.~{Coudreau}, N.~{Hallet}, and J.~M. {Martin}.
\newblock {Kinematics of the Local Universe. XIII. 21-cm line measurements of
  452 galaxies with the Nan{\c{c}}ay radiotelescope, JHK Tully-Fisher relation,
  and preliminary maps of the peculiar velocity field}.
\newblock \emph{\aap}, 465\penalty0 (1):\penalty0 71--85, April 2007.
\newblock \doi{10.1051/0004-6361:20066187}.

\bibitem[{Tully} et~al.(2013){Tully}, {Courtois}, {Dolphin}, {Fisher},
  {H{\'e}raudeau}, {Jacobs}, {Karachentsev}, {Makarov}, {Makarova},
  {Mitronova}, {Rizzi}, {Shaya}, {Sorce}, and {Wu}]{tully}
R.~Brent {Tully}, H{\'e}l{\`e}ne~M. {Courtois}, Andrew~E. {Dolphin}, J.~Richard
  {Fisher}, Philippe {H{\'e}raudeau}, Bradley~A. {Jacobs}, Igor~D.
  {Karachentsev}, Dmitry {Makarov}, Lidia {Makarova}, Sofia {Mitronova}, Luca
  {Rizzi}, Edward~J. {Shaya}, Jenny~G. {Sorce}, and Po-Feng {Wu}.
\newblock {Cosmicflows-2: The Data}.
\newblock \emph{\aj}, 146\penalty0 (4):\penalty0 86, October 2013.
\newblock \doi{10.1088/0004-6256/146/4/86}.

\bibitem[{Saulder} et~al.(2016){Saulder}, {van Kampen}, {Chilingarian},
  {Mieske}, and {Zeilinger}]{saulder}
Christoph {Saulder}, Eelco {van Kampen}, Igor~V. {Chilingarian}, Steffen
  {Mieske}, and Werner~W. {Zeilinger}.
\newblock {The matter distribution in the local Universe as derived from galaxy
  groups in SDSS DR12 and 2MRS}.
\newblock \emph{\aap}, 596:\penalty0 A14, November 2016.
\newblock \doi{10.1051/0004-6361/201526711}.

\bibitem[{Ma} et~al.(2014){Ma}, {Greene}, {McConnell}, {Janish}, {Blakeslee},
  {Thomas}, and {Murphy}]{ma2014}
Chung-Pei {Ma}, Jenny~E. {Greene}, Nicholas {McConnell}, Ryan {Janish}, John~P.
  {Blakeslee}, Jens {Thomas}, and Jeremy~D. {Murphy}.
\newblock {The MASSIVE Survey. I. A Volume-limited Integral-field Spectroscopic
  Study of the Most Massive Early-type Galaxies within 108 Mpc}.
\newblock \emph{\apj}, 795\penalty0 (2):\penalty0 158, November 2014.
\newblock \doi{10.1088/0004-637X/795/2/158}.

\bibitem[{Crook} et~al.(2007){Crook}, {Huchra}, {Martimbeau}, {Masters},
  {Jarrett}, and {Macri}]{crook}
Aidan~C. {Crook}, John~P. {Huchra}, Nathalie {Martimbeau}, Karen~L. {Masters},
  Tom {Jarrett}, and Lucas~M. {Macri}.
\newblock {Groups of Galaxies in the Two Micron All Sky Redshift Survey}.
\newblock \emph{\apj}, 655\penalty0 (2):\penalty0 790--813, February 2007.
\newblock \doi{10.1086/510201}.

\bibitem[{Tonry} et~al.(2001){Tonry}, {Dressler}, {Blakeslee}, {Ajhar},
  {Fletcher}, {Luppino}, {Metzger}, and {Moore}]{tonry}
John~L. {Tonry}, Alan {Dressler}, John~P. {Blakeslee}, Edward~A. {Ajhar},
  Andr{\'e}~B. {Fletcher}, Gerard~A. {Luppino}, Mark~R. {Metzger}, and
  Christopher~B. {Moore}.
\newblock {The SBF Survey of Galaxy Distances. IV. SBF Magnitudes, Colors, and
  Distances}.
\newblock \emph{\apj}, 546\penalty0 (2):\penalty0 681--693, January 2001.
\newblock \doi{10.1086/318301}.

\bibitem[{Snowden} and {Kuntz}(2011)]{ESAS_AAS}
Steven~L. {Snowden} and K.~D. {Kuntz}.
\newblock {Analysis of XMM-Newton Data from Extended Sources and the Diffuse
  X-ray Background}.
\newblock In \emph{American Astronomical Society Meeting Abstracts \#217},
  volume 217 of \emph{American Astronomical Society Meeting Abstracts}, page
  344.17, January 2011.

\bibitem[{Fruscione} et~al.(2006){Fruscione}, {McDowell}, {Allen},
  {Brickhouse}, {Burke}, {Davis}, {Durham}, {Elvis}, {Galle}, {Harris},
  {Huenemoerder}, {Houck}, {Ishibashi}, {Karovska}, {Nicastro}, {Noble},
  {Nowak}, {Primini}, {Siemiginowska}, {Smith}, and {Wise}]{CIAO}
Antonella {Fruscione}, Jonathan~C. {McDowell}, Glenn~E. {Allen}, Nancy~S.
  {Brickhouse}, Douglas~J. {Burke}, John~E. {Davis}, Nick {Durham}, Martin
  {Elvis}, Elizabeth~C. {Galle}, Daniel~E. {Harris}, David~P. {Huenemoerder},
  John~C. {Houck}, Bish {Ishibashi}, Margarita {Karovska}, Fabrizio {Nicastro},
  Michael~S. {Noble}, Michael~A. {Nowak}, Frank~A. {Primini}, Aneta
  {Siemiginowska}, Randall~K. {Smith}, and Michael {Wise}.
\newblock \emph{{CIAO: Chandra's data analysis system}}, volume 6270 of
  \emph{Society of Photo-Optical Instrumentation Engineers (SPIE) Conference
  Series}, page 62701V.
\newblock 2006.
\newblock \doi{10.1117/12.671760}.

\bibitem[{Trumpler} and {Weaver}(1953)]{trumpler}
Robert~J. {Trumpler} and Harold~F. {Weaver}.
\newblock \emph{{Statistical astronomy}}.
\newblock University of California Press, Berkeley and Los Angeles, 1953.

\bibitem[{McMillan} et~al.(1989){McMillan}, {Kowalski}, and {Ulmer}]{McMillan}
S.~L.~W. {McMillan}, M.~P. {Kowalski}, and M.~P. {Ulmer}.
\newblock {X-Ray Morphologies of Abell Clusters}.
\newblock \emph{\apjs}, 70:\penalty0 723, August 1989.
\newblock \doi{10.1086/191356}.

\bibitem[{Carter} and {Metcalfe}(1980)]{carterMetcalfe}
D.~{Carter} and N.~{Metcalfe}.
\newblock {The morphology of clusters of galaxies.}
\newblock \emph{\mnras}, 191:\penalty0 325--337, May 1980.
\newblock \doi{10.1093/mnras/191.2.325}.

\bibitem[Nelder and Mead(1965)]{NelderMead}
J.~A. Nelder and R.~Mead.
\newblock A simplex method for function minimization.
\newblock \emph{The Computer Journal}, 7\penalty0 (4):\penalty0 308--313,
  January 1965.
\newblock \doi{10.1093/comjnl/7.4.308}.
\newblock URL \url{https://doi.org/10.1093/comjnl/7.4.308}.

\bibitem[Press(2007)]{numericalRecipes}
William~H. Press.
\newblock \emph{Numerical Recipes 3rd Edition: The Art of Scientific
  Computing}.
\newblock Cambridge University Press, sep 2007.
\newblock ISBN 0521880688.
\newblock URL \url{https://www.xarg.org/ref/a/0521880688/}.

\bibitem[Gao and Han(2010)]{Gao2010}
Fuchang Gao and Lixing Han.
\newblock Implementing the nelder-mead simplex algorithm with~adaptive
  parameters.
\newblock \emph{Computational Optimization and Applications}, 51\penalty0
  (1):\penalty0 259--277, May 2010.
\newblock \doi{10.1007/s10589-010-9329-3}.
\newblock URL \url{https://doi.org/10.1007/s10589-010-9329-3}.

\bibitem[{Buote} and {Canizares}(1996{\natexlab{b}})]{buoteTwist}
David~A. {Buote} and Claude~R. {Canizares}.
\newblock {The Twisting X-Ray Isophotes of the Elliptical Galaxy NGC 720}.
\newblock \emph{\apj}, 468:\penalty0 184, September 1996{\natexlab{b}}.
\newblock \doi{10.1086/177680}.

\bibitem[{Romanowsky} and {Kochanek}(1998)]{romanowsky}
Aaron~J. {Romanowsky} and Christopher~S. {Kochanek}.
\newblock {Twisting of X-Ray Isophotes in Triaxial Galaxies}.
\newblock \emph{\apj}, 493\penalty0 (2):\penalty0 641--649, January 1998.
\newblock \doi{10.1086/305151}.

\bibitem[{Memola} et~al.(2011){Memola}, {Salucci}, and {Babi{\'c}}]{7785_mass}
E.~{Memola}, P.~{Salucci}, and A.~{Babi{\'c}}.
\newblock {Dark matter halos around isolated ellipticals}.
\newblock \emph{\aap}, 534:\penalty0 A50, October 2011.
\newblock \doi{10.1051/0004-6361/201015667}.

\bibitem[{Buote}(2017)]{6482_mass}
David~A. {Buote}.
\newblock {The Unusually High Halo Concentration of the Fossil Group NGC 6482:
  Evidence for Weak Adiabatic Contraction}.
\newblock \emph{\apj}, 834\penalty0 (2):\penalty0 164, January 2017.
\newblock \doi{10.3847/1538-4357/834/2/164}.

\bibitem[{Humphrey} et~al.(2011){Humphrey}, {Buote}, {Canizares}, {Fabian}, and
  {Miller}]{720_mass}
Philip~J. {Humphrey}, David~A. {Buote}, Claude~R. {Canizares}, Andrew~C.
  {Fabian}, and Jon~M. {Miller}.
\newblock {A Census of Baryons and Dark Matter in an Isolated, Milky Way Sized
  Elliptical Galaxy}.
\newblock \emph{\apj}, 729\penalty0 (1):\penalty0 53, March 2011.
\newblock \doi{10.1088/0004-637X/729/1/53}.

\bibitem[{Humphrey} et~al.(2012){Humphrey}, {Buote}, {O'Sullivan}, and
  {Ponman}]{elixr_1521}
Philip~J. {Humphrey}, David~A. {Buote}, Ewan {O'Sullivan}, and Trevor~J.
  {Ponman}.
\newblock {The ElIXr Galaxy Survey. II. Baryons and Dark Matter in an Isolated
  Elliptical Galaxy}.
\newblock \emph{\apj}, 755\penalty0 (2):\penalty0 166, August 2012.
\newblock \doi{10.1088/0004-637X/755/2/166}.

\bibitem[{O'Sullivan} and {Ponman}(2004)]{ngc4555_mass}
E.~{O'Sullivan} and T.~J. {Ponman}.
\newblock {The isolated elliptical NGC 4555 observed with Chandra}.
\newblock \emph{\mnras}, 354\penalty0 (3):\penalty0 935--944, November 2004.
\newblock \doi{10.1111/j.1365-2966.2004.08257.x}.

\bibitem[{Brighenti} and {Mathews}(1997)]{brighenti}
Fabrizio {Brighenti} and William~G. {Mathews}.
\newblock {X-Ray Observations and the Structure of Elliptical Galaxies}.
\newblock \emph{\apjl}, 486\penalty0 (2):\penalty0 L83--L86, September 1997.
\newblock \doi{10.1086/310857}.

\bibitem[{Gerhard} et~al.(2001){Gerhard}, {Kronawitter}, {Saglia}, and
  {Bender}]{gerhard}
Ortwin {Gerhard}, Andi {Kronawitter}, R.~P. {Saglia}, and Ralf {Bender}.
\newblock {Dynamical Family Properties and Dark Halo Scaling Relations of Giant
  Elliptical Galaxies}.
\newblock \emph{\aj}, 121\penalty0 (4):\penalty0 1936--1951, April 2001.
\newblock \doi{10.1086/319940}.

\end{thebibliography}

\appendix
\setcounter{table}{0}
\renewcommand\thetable{\Alph{section}.\arabic{table}}

\section{Results of Model Fitting Procedure}
\begin{table*}[tbph!]
\centering
\def\arraystretch{1.25}
\setlength{\tabcolsep}{8pt}

\begin{tabular}{lc|cccc|cccc}
	\hline
\multicolumn{10}{c}{XMM}\\
	\hline\hline
	&&\multicolumn{4}{c}{Prolate}&\multicolumn{4}{c}{Oblate}\\
	\cline{3-10}
	Galaxy & Profile & $a_s$ ('') & $\Gamma$ &$\epsilon$&$\chi_{red}^2$&$a_s$ ('') &$\Gamma$&$\epsilon$&$\chi_{red}^2$\\
	\hline
\multirow{3}{*}{IC4451}&  NFW &61.0 & 7.8 & 0.34 & 8.3&24.5 & 7.9 & 0.35 & 8.2 \\  
& Hernq. &126.2 & 7.0 & 0.31 & 9.8&45.8 & 6.8 & 0.34 & 9.6 \\  
& pIso &10.2 & 6.6 & 0.32 & 10.2&4.8 & 7.4 & 0.26 & 14.1 \\  
\hline
\multirow{3}{*}{IC4956}&  NFW &40.7 & 8.4 & 0.17 & 0.9&15.8 & 8.4 & 0.19 & 0.9 \\  
& Hernq. &85.0 & 7.8 & 0.16 & 1.0&34.2 & 7.8 & 0.16 & 1.1 \\  
& pIso &6.1 & 7.0 & 0.15 & 0.9&2.4 & 7.4 & 0.12 & 1.2 \\  
\hline
\multirow{3}{*}{NGC1521}&  NFW &56.9 & 7.9 & 0.43 & 13.1&21.7 & 8.0 & 0.44 & 12.7 \\  
& Hernq. &104.6 & 6.8 & 0.44 & 14.3&40.8 & 6.9 & 0.46 & 13.7 \\  
& pIso &8.7 & 6.6 & 0.42 & 14.0&2.7 & 6.9 & 0.49 & 13.6 \\  
\hline
\multirow{3}{*}{NGC4125}&  NFW &77.0 & 7.4 & 0.48 & 1.3&29.0 & 7.6 & 0.5 & 1.2 \\  
& Hernq. &123.9 & 6.2 & 0.48 & 1.3&49.6 & 6.3 & 0.51 & 1.3 \\  
& pIso &11.2 & 6.2 & 0.48 & 1.3&4.2 & 6.3 & 0.52 & 1.2 \\  
\hline
\multirow{3}{*}{NGC4555}&  NFW &24.2 & 6.9 & 0.28 & 5.2&8.0 & 6.9 & 0.28 & 4.9 \\  
& Hernq. &58.2 & 6.2 & 0.36 & 7.6&19.5 & 6.2 & 0.32 & 7.0 \\  
& pIso &1.8 & 7.6 & 0.27 & 2.3&0.6 & 7.7 & 0.31 & 2.2 \\  
\hline
\multirow{3}{*}{NGC6482}&  NFW &46.3 & 8.0 & 0.21 & 7.9&16.6 & 7.9 & 0.23 & 7.8 \\  
& Hernq. &102.9 & 7.2 & 0.22 & 10.6&40.0 & 7.2 & 0.26 & 11.0 \\  
& pIso &5.6 & 7.6 & 0.2 & 6.6&2.2 & 7.6 & 0.23 & 6.4 \\  
\hline
\multirow{3}{*}{NGC7785}&  NFW &52.0 & 7.6 & 0.23 & 4.4&20.6 & 7.8 & 0.25 & 4.2 \\  
& Hernq. &194.4 & 9.8 & 0.17 & 2.0&47.9 & 7.5 & 0.18 & 3.1 \\  
& pIso &29.8 & 9.7 & 0.16 & 1.2&7.3 & 7.8 & 0.21 & 1.8 \\  
\hline
\multirow{3}{*}{NGC7796}&  NFW &47.9 & 7.3 & 0.45 & 3.0&22.9 & 8.0 & 0.43 & 3.6 \\  
& Hernq. &107.4 & 6.7 & 0.42 & 3.6&40.9 & 6.6 & 0.44 & 3.6 \\  
& pIso &4.8 & 6.6 & 0.42 & 2.7&5.9 & 7.6 & 0.32 & 7.4 \\  
\hline
\multirow{3}{*}{NGC953}&  NFW &36.8 & 8.2 & 0.27 & 3.2&11.3 & 7.7 & 0.24 & 2.9 \\  
& Hernq. &73.4 & 7.6 & 0.22 & 3.9&28.0 & 7.6 & 0.29 & 4.0 \\  
& pIso &3.8 & 7.4 & 0.25 & 2.9&1.4 & 7.4 & 0.24 & 2.8 \\  
\hline
\multirow{3}{*}{NGC720}&  NFW &79.1 & 7.6 & 0.34 & 8.9&30.7 & 7.6 & 0.35 & 8.4 \\  
& Hernq. &126.7 & 6.4 & 0.26 & 12.2&60.0 & 6.7 & 0.33 & 9.5 \\  
& pIso &13.2 & 6.5 & 0.3 & 10.7&4.6 & 6.6 & 0.34 & 9.9 \\  

\hline\hline
\end{tabular}
\caption[XMM Fit Parameters]{Results of the best-fitting parameters for the XMM data. We show the results for both the prolate and oblate configurations for each mass profile. The reduced-$\chi^2$ is also provided and all fits have 37 degrees of freedom. \label{tab:xmm_fits} }
\end{table*}
\newpage
\begin{table*}[tbph!]
\centering
\def\arraystretch{1.25}
\setlength{\tabcolsep}{8pt}

\begin{tabular}{lc|cccc|cccc}
%\def\arraystretch{1.65}
%setlength{\tabcolsep}{10pt}

	\hline
\multicolumn{10}{c}{Chandra}\\
	\hline\hline
	&&\multicolumn{4}{c}{Prolate}&\multicolumn{4}{c}{Oblate}\\
	\cline{3-10}
	Galaxy & Profile & $a_s$ ('') & $\Gamma$ &$\epsilon$&$\chi_{red}^2$&$a_s$ ('') &$\Gamma$&$\epsilon$&$\chi_{red}^2$\\
	\hline
	
\multirow{3}{*}{IC4451}&  NFW &12.4 & 6.3 & 0.46 & 4.1&11.8 & 6.3 & 0.44 & 4.0 \\  
& Hernq. &27.0 & 5.8 & 0.21 & 7.7&25.1 & 5.7 & 0.51 & 4.6 \\  
& pIso &0.8 & 7.0 & 0.36 & 3.4&0.9 & 6.8 & 0.37 & 3.3 \\  
\hline
\multirow{3}{*}{NGC1521}&  NFW &13.8 & 6.2 & 0.41 & 2.8&13.2 & 6.1 & 0.41 & 2.9 \\  
& Hernq. &26.4 & 5.6 & 0.19 & 6.1&27.0 & 5.5 & 0.26 & 5.0 \\  
& pIso &1.1 & 6.5 & 0.29 & 3.2&1.4 & 6.3 & 0.45 & 2.5 \\  
\hline
\multirow{3}{*}{NGC4125}&  NFW &43.4 & 6.8 & 0.48 & 5.8&32.4 & 6.5 & 0.59 & 4.3 \\  
& Hernq. &166.3 & 7.4 & 0.5 & 12.0&68.0 & 5.8 & 0.6 & 4.9 \\  
& pIso &9.9 & 5.8 & 0.5 & 7.2&4.5 & 6.0 & 0.41 & 9.2 \\  
\hline
\multirow{3}{*}{NGC4555}&  NFW &7.2 & 7.5 & 0.36 & 1.4&7.2 & 7.7 & 0.24 & 1.1 \\  
& Hernq. &16.1 & 7.0 & 0.25 & 1.6&18.2 & 7.1 & 0.51 & 2.7 \\  
& pIso &0.8 & 8.4 & 0.41 & 1.8&0.5 & 8.7 & 0.38 & 1.1 \\  
\hline
\multirow{3}{*}{NGC57}&  NFW &12.8 & 8.7 & 0.34 & 1.6&8.7 & 7.7 & 0.32 & 1.3 \\  
& Hernq. &17.7 & 6.9 & 0.28 & 1.4&15.5 & 6.4 & 0.46 & 1.6 \\  
& pIso &1.5 & 7.5 & 0.34 & 1.2&1.3 & 7.6 & 0.34 & 1.3 \\  
\hline
\multirow{3}{*}{NGC6482}&  NFW &23.8 & 8.2 & 0.2 & 43.5&23.3 & 8.1 & 0.22 & 42.6 \\  
& Hernq. &51.8 & 7.2 & 0.32 & 61.0&47.0 & 7.3 & 0.22 & 53.7 \\  
& pIso &2.6 & 8.2 & 0.2 & 28.7&2.2 & 8.4 & 0.25 & 29.7 \\  
\hline
\multirow{3}{*}{NGC7796}&  NFW &9.0 & 6.8 & 0.41 & 4.7&8.3 & 6.7 & 0.44 & 4.4 \\  
& Hernq. &18.0 & 6.2 & 0.22 & 7.8&19.5 & 6.1 & 0.43 & 5.9 \\  
& pIso &0.5 & 7.6 & 0.38 & 2.6&0.4 & 7.8 & 0.41 & 2.3 \\  
\hline
\multirow{3}{*}{NGC953}&  NFW &6.1 & 7.0 & 0.3 & 1.9&6.2 & 7.1 & 0.34 & 1.9 \\  
& Hernq. &13.5 & 6.5 & 0.21 & 2.4&13.3 & 6.3 & 0.35 & 2.2 \\  
& pIso &0.5 & 7.6 & 0.27 & 1.6&0.7 & 7.5 & 0.36 & 1.6 \\  
\hline
\multirow{3}{*}{NGC720}&  NFW &46.8 & 7.3 & 0.34 & 7.2&45.5 & 7.3 & 0.38 & 7.0 \\  
& Hernq. &94.2 & 6.4 & 0.35 & 8.2&88.8 & 6.4 & 0.36 & 8.1 \\  
& pIso &5.9 & 6.5 & 0.34 & 7.3&5.6 & 6.5 & 0.35 & 6.8 \\  
\hline\hline
\end{tabular}
\caption[Chandra Fit Parameters]{\label{tab:chandra_fits}Results of the best-fitting parameters for the Chandra data. We show the results for both the prolate and oblate configurations for each mass profile. The reduced-$\chi^2$ is also provided and all fits have 37 degrees of freedom.}
\end{table*}

\begin{table*}[tbph!]
\centering
\def\arraystretch{1.2}
\setlength{\tabcolsep}{8pt}

\begin{tabular}{lc|cccc|cccc}
	\hline
\multicolumn{10}{c}{XMM}\\
	\hline\hline
	&&\multicolumn{4}{c}{Prolate}&\multicolumn{4}{c}{Oblate}\\
	\cline{3-10}
	Galaxy & Profile & $a_s$ ('') & $\Gamma$ &$\epsilon$&$\chi_{red}^2$&$a_s$ ('') &$\Gamma$&$\epsilon$&$\chi_{red}^2$\\
	\hline
\multirow{3}{*}{IC4451}&  NFW &62.9 & 7.8 & 0.26 & 10.0&24.1 & 8.0 & 0.32 & 13.8 \\  
& Hernq. &125.8 & 7.1 & 0.22 & 11.0&49.2 & 7.1 & 0.25 & 11.2 \\  
& pIso &8.3 & 6.7 & 0.3 & 12.1&2.7 & 6.8 & 0.38 & 14.9 \\  
\hline
\multirow{3}{*}{IC4956}&  NFW &35.4 & 8.0 & 0.17 & 0.8&12.5 & 7.6 & 0.19 & 0.9 \\  
& Hernq. &96.7 & 8.2 & 0.16 & 1.2&35.2 & 8.0 & 0.17 & 1.1 \\  
& pIso &8.5 & 7.4 & 0.28 & 2.5&2.2 & 6.9 & 0.19 & 1.0 \\  
\hline
\multirow{3}{*}{NGC1521}&  NFW &42.7 & 7.6 & 0.26 & 4.1&21.2 & 8.0 & 0.4 & 14.8 \\  
& Hernq. &111.7 & 7.3 & 0.24 & 5.8&33.4 & 6.8 & 0.24 & 4.5 \\  
& pIso &7.4 & 6.8 & 0.32 & 5.7&2.9 & 6.8 & 0.44 & 15.4 \\  
\hline
\multirow{3}{*}{NGC4125}&  NFW &54.6 & 7.0 & 0.44 & 1.4&27.3 & 7.3 & 0.52 & 1.4 \\  
& Hernq. &121.2 & 6.3 & 0.46 & 1.4&48.6 & 6.3 & 0.49 & 1.3 \\  
& pIso &9.5 & 6.2 & 0.37 & 1.6&3.9 & 6.2 & 0.51 & 1.4 \\  
\hline
\multirow{3}{*}{NGC4555}&  NFW &22.8 & 6.9 & 0.33 & 6.9&8.3 & 6.9 & 0.32 & 8.6 \\  
& Hernq. &74.3 & 6.2 & 0.55 & 14.3&21.2 & 6.2 & 0.39 & 10.1 \\  
& pIso &1.8 & 7.5 & 0.29 & 2.7&0.6 & 7.7 & 0.32 & 4.1 \\  
\hline
\multirow{3}{*}{NGC57}&  NFW &24.4 & 8.1 & 0.24 & 1.1&12.1 & 8.7 & 0.37 & 3.1 \\  
& Hernq. &48.0 & 7.2 & 0.18 & 1.0&19.6 & 7.3 & 0.21 & 1.0 \\  
& pIso &3.9 & 8.0 & 0.25 & 1.2&1.4 & 8.1 & 0.2 & 1.1 \\  
\hline
\multirow{3}{*}{NGC6482}&  NFW &51.4 & 8.4 & 0.29 & 13.6&19.4 & 8.5 & 0.17 & 15.4 \\  
& Hernq. &93.0 & 7.4 & 0.08 & 13.6&40.0 & 7.6 & 0.2 & 13.9 \\  
& pIso &7.6 & 7.9 & 0.25 & 11.7&2.7 & 8.0 & 0.16 & 14.4 \\  
\hline
\multirow{3}{*}{NGC7785}&  NFW &59.1 & 8.1 & 0.23 & 4.7&36.0 & 10.0 & 0.28 & 3.9 \\  
& Hernq. &120.3 & 7.4 & 0.21 & 4.0&76.2 & 9.8 & 0.2 & 2.6 \\  
& pIso &23.6 & 8.4 & 0.18 & 1.8&11.6 & 9.7 & 0.18 & 1.7 \\  
\hline
\multirow{3}{*}{NGC7796}&  NFW &44.4 & 7.2 & 0.41 & 3.5&13.5 & 7.0 & 0.48 & 3.9 \\  
& Hernq. &96.9 & 6.5 & 0.42 & 4.3&28.9 & 6.2 & 0.38 & 3.5 \\  
& pIso &4.4 & 6.6 & 0.32 & 2.9&1.8 & 6.6 & 0.48 & 3.5 \\  
\hline
\multirow{3}{*}{NGC953}&  NFW &27.4 & 7.7 & 0.22 & 2.9&12.5 & 7.9 & 0.28 & 4.8 \\  
& Hernq. &92.4 & 8.0 & 0.26 & 6.5&22.3 & 6.9 & 0.29 & 4.3 \\  
& pIso &4.3 & 7.5 & 0.27 & 3.6&1.6 & 7.5 & 0.27 & 4.5 \\  
\hline
\multirow{3}{*}{NGC720}&  NFW &75.6 & 7.7 & 0.23 & 11.8&32.4 & 7.8 & 0.32 & 12.0 \\  
& Hernq. &143.1 & 6.7 & 0.2 & 13.8&53.4 & 6.6 & 0.25 & 13.3 \\  
& pIso &13.4 & 6.5 & 0.33 & 15.2&4.7 & 6.6 & 0.33 & 14.3 \\  
\hline\hline
\end{tabular}
\caption[XMM Fit Parameters]{Results of the best-fitting parameters for the XMM data in the range restricted scenario. We show the results for both the prolate and oblate configurations for each mass profile. \label{tab:xmm_fits_range} }
\end{table*}
\newpage
\begin{table*}[tbph!]
\centering
\def\arraystretch{1.25}
\setlength{\tabcolsep}{8pt}

\begin{tabular}{lc|cccc|cccc}
%\def\arraystretch{1.65}
%setlength{\tabcolsep}{10pt}

	\hline
\multicolumn{10}{c}{Chandra}\\
	\hline\hline
	&&\multicolumn{4}{c}{Prolate}&\multicolumn{4}{c}{Oblate}\\
	\cline{3-10}
	Galaxy & Profile & $a_s$ ('') & $\Gamma$ &$\epsilon$&$\chi_{red}^2$&$a_s$ ('') &$\Gamma$&$\epsilon$&$\chi_{red}^2$\\
	\hline
	\multirow{3}{*}{IC4451}&  NFW &12.1 & 6.6 & 0.43 & 4.3&11.5 & 6.5 & 0.47 & 4.5 \\  
& Hernq. &32.7 & 5.8 & 0.22 & 6.1&26.2 & 5.8 & 0.42 & 4.6 \\  
& pIso &1.0 & 6.9 & 0.36 & 3.3&1.1 & 6.7 & 0.45 & 3.3 \\  
\hline
\multirow{3}{*}{NGC1521}&  NFW &19.8 & 7.0 & 0.37 & 2.8&18.3 & 6.9 & 0.4 & 3.7 \\  
& Hernq. &40.1 & 6.1 & 0.22 & 3.4&35.3 & 6.0 & 0.35 & 3.1 \\  
& pIso &2.5 & 6.6 & 0.26 & 2.4&1.9 & 6.6 & 0.37 & 3.2 \\  
\hline
\multirow{3}{*}{NGC4125}&  NFW &41.2 & 7.0 & 0.52 & 5.4&39.4 & 7.0 & 0.56 & 5.4 \\  
& Hernq. &80.5 & 6.1 & 0.52 & 6.3&78.5 & 6.1 & 0.56 & 5.9 \\  
& pIso &5.5 & 6.1 & 0.5 & 6.4&4.9 & 6.1 & 0.58 & 5.5 \\  
\hline
\multirow{3}{*}{NGC4555}&  NFW &8.0 & 7.8 & 0.12 & 2.9&8.3 & 8.0 & 0.33 & 2.3 \\  
& Hernq. &15.4 & 7.3 & 0.05 & 3.6&18.0 & 7.2 & 0.36 & 2.6 \\  
& pIso &0.7 & 8.7 & 0.29 & 1.4&0.8 & 8.8 & 0.28 & 1.7 \\  
\hline
\multirow{3}{*}{NGC57}&  NFW &6.4 & 6.8 & 0.19 & 2.5&6.8 & 7.2 & 0.14 & 2.7 \\  
& Hernq. &26.9 & 8.2 & 0.22 & 2.2&15.7 & 6.7 & 0.36 & 1.2 \\  
& pIso &1.5 & 7.1 & 0.37 & 1.5&1.2 & 7.4 & 0.36 & 1.4 \\  
\hline
\multirow{3}{*}{NGC6482}&  NFW &19.9 & 8.0 & 0.01 & 62.3&21.3 & 8.0 & 0.17 & 58.8 \\  
& Hernq. &44.7 & 7.2 & 0.09 & 77.8&44.8 & 7.2 & 0.16 & 75.7 \\  
& pIso &3.7 & 7.7 & 0.4 & 67.7&2.3 & 8.3 & 0.17 & 33.3 \\  
\hline
\multirow{3}{*}{NGC7796}&  NFW &9.3 & 6.6 & 0.34 & 6.8&8.7 & 6.7 & 0.53 & 6.0 \\  
& Hernq. &16.0 & 6.1 & 0.05 & 11.8&20.0 & 6.0 & 0.54 & 7.0 \\  
& pIso &0.4 & 7.8 & 0.36 & 3.5&0.4 & 7.7 & 0.5 & 3.2 \\  
\hline
\multirow{3}{*}{NGC953}&  NFW &6.4 & 6.9 & 0.24 & 3.0&6.7 & 7.3 & 0.13 & 3.0 \\  
& Hernq. &17.4 & 7.0 & 0.27 & 2.5&16.2 & 6.8 & 0.31 & 2.4 \\  
& pIso &1.0 & 7.9 & 0.3 & 2.0&0.8 & 7.8 & 0.35 & 2.0 \\  
\hline
\multirow{3}{*}{NGC720}&  NFW &20.7 & 6.7 & 0.53 & 15.7&31.2 & 6.4 & 0.34 & 3.1 \\  
& Hernq. &60.7 & 5.9 & 0.41 & 5.6&65.3 & 5.8 & 0.3 & 3.5 \\  
& pIso &4.0 & 6.2 & 0.27 & 2.1&3.2 & 6.4 & 0.35 & 2.4 \\  
\hline
\hline
\end{tabular}
\caption[Chandra Fit Parameters]{\label{tab:chandra_fits_range} Results of the best-fitting parameters for the Chandra data in the range restricted scenario. We show the results for both the prolate and oblate configurations for each mass profile. }
\end{table*}
\end{document}